\documentclass{emulateapj}
\usepackage{epsfig}
\usepackage{color}
\newcommand{\msun}{\,\rm M_\odot}
\newcommand{\etal}{et al.\ }
\newcommand{\kmsmpc}{\,{\rm km\,s^{-1}\,Mpc^{-1}}}
\newcommand{\be}{\begin{equation}}
\newcommand{\ee}{\end{equation}}
\newcommand{\ba}{\begin{eqnarray}}
\newcommand{\ea}{\end{eqnarray}}

\newcommand{\mhost}{M_{\rm host}}
\newcommand{\rtwo}{r_{200}}

\newcommand{\msub}{M_{\rm sub}}

\newcommand{\vmax}{V_{\rm max}}
\newcommand{\peakvmax}{V_{\rm max,p}}
\newcommand{\rvhost}{r_{\rm Vhost}}
\newcommand{\vhost}{V_{\rm host}}

\newcommand{\kms}{\,{\rm km\,s^{-1}}}
\newcommand{\sden}{\,{\rm M_\odot\,pc^{-3}}}

\def\spose#1{\hbox to 0pt{#1\hss}}
\def\lta{\mathrel{\spose{\lower 3pt\hbox{$\mathchar"218$}} \raise 2.0pt\hbox{$\mathchar"13C$}}}
\def\gta{\mathrel{\spose{\lower 3pt\hbox{$\mathchar"218$}} \raise 2.0pt\hbox{$\mathchar"13E$}}}

\lefthead{Madau et al.}
\righthead{Dark matter subhalos and dwarf satellites}



\begin{document}
\submitted{}
\accepted{February 14, 2008}

\title{Dark matter subhalos and the dwarf satellites of the Milky Way}

\author{Piero Madau\altaffilmark{1}, J\"urg Diemand\altaffilmark{1,2}, 
\& Michael Kuhlen\altaffilmark{3}}

\altaffiltext{1}{Department of Astronomy \& Astrophysics, University of California, 
Santa Cruz, CA 95064.}
\altaffiltext{2}{Hubble Fellow.}
\altaffiltext{3}{Institute for Advanced Study, Einstein Drive, Princeton, NJ 08540.}
\email{diemand@ucolick.org, mqk@ias.edu, pmadau@ucolick.org.}

\begin{abstract}
  The Via Lactea simulation of the cold dark matter halo of the Milky
  Way predicts the existence of many thousands of bound subhalos with
  masses above a few $\times 10^6\,\msun$, distributed approximately
  with equal mass per decade of mass. Here we show that: a) a similar
  steeply rising subhalo mass function is also present at redshift
  $z=0.5$ in an elliptical-sized halo simulated with comparable
  resolution in a different cosmology. Compared to Via Lactea, this
  run produces nearly a factor of two more subhalos with large
  circular velocities; b) the fraction of Via Lactea mass brought in
  by subhalos that have a surviving bound remnant today within $\rtwo$
  and with present-day peak circular velocity $\vmax>2\,\kms$
  ($>10\,\kms$) is 45\% (30\%). Most of the Via Lactea mass is acquired in 
  resolved discrete clumps, with no evidence for a significant smooth infall; 
  c) because of tidal mass loss, the
  number of subhalos surviving today that reached a peak circular
  velocity of $>10\,\kms$ throughout their lifetime exceeds half a
  thousand, five times larger than their present-day abundance and
  more than twenty times larger than the number of known satellites of
  the Milky Way; e) unless the circular velocity profiles of Galactic
  satellites peak at values significantly higher that expected from
  the stellar line-of-sight velocity dispersion, only about one in
  five subhalos with $\vmax>20\,\kms$ today must be housing a luminous
  dwarf. Any mechanism that suppresses star formation in small
  subhalos must start acting early, at redshift $z>3$; f) nearly 600
  halos with masses greater than $10^7\,\msun$ are found today in the
  ``field'' between $\rtwo$ and $1.5\rtwo$, i.e. small dark matter
  clumps appear to be relatively inefficient at forming stars even
  well beyond the virial radius; g) the observed Milky Way satellites
  appear to follow the overall dark matter distribution of Via Lactea,
  while the largest simulated subhalos today are found preferentially
  at larger radii; h) subhalos have central densities that increase
  with $\vmax$ and reach $\rho_{\rm DM}=0.1-0.3\,\sden$, comparable to
  the central densities inferred in dwarf spheroidals with core radii
  $>250$ pc.
\end{abstract}

 
\section{Introduction}

In the standard cosmological paradigm of structure formation
($\Lambda$CDM), the universe is dominated by cold, collisionless dark
matter, and objects like the halo of our Milky Way are assembled via
the hierarchical merging and accretion of smaller progenitors.
Subunits collapse at high redshift, and when they merge into larger
hosts their high densities allow them to resist the strong tidal
forces that act to destroy them.  Indeed, numerical simulations have
shown that CDM halos are lumpy, teeming with self-bound subhalos on
all resolved mass scales. It is now well established that the
predicted subhalo counts vastly exceed the number of observed
satellites of the Milky Way \citep{Moore1999a,Klypin1999}, a ``missing
satellite'' or ``substructure'' problem that has been the subject of
many recent studies.  While some models have attempted to solve the
apparent small-scale difficulties of CDM at a more fundamental level,
i.e. by reducing small-scale power (e.g.
\citealt{Kamionkowski2000,Colin2000}), astrophysical solutions that
quench gas accretion and star formation in small halos have also been
proposed as a plausible way out. The latter category includes cosmic
reionization (e.g.
\citealt{Bullock2000,Benson2002,Somerville2002,Kravtsov2004,Moore2006}),
supernova feedback (e.g. \citealt{Dekel86,MacLow1999,Mori2002}), and
gas stripping by ram pressure (e.g. \citealt{Mayer06}) as favourite
suppression mechanisms. Despite a wealth of ideas, however, a full
characterization of the substructure problem is limited by three main
uncertainties: 1) the old tally of Milky Way luminous satellites is
very incomplete, and is being rapidly revised (e.g.
\citealt{Belokurov2007}) by the discovery of a new, large population
of ultra-faint dwarfs in the Sloan Digital Sky Survey (SDSS); 2) in
cosmological $N$-body simulations, the ability of subhalos to survive
the hierarchical clustering process as substructure within the host is
particularly sensitive to resolution issues, as subhalos with
numerically softened central densities are more easily disrupted by
tidal forces (e.g. \citealt{Moore1996,Diemand2007a}); and 3) a
detailed comparison between theory and observation in terms of, e.g.,
the peak circular velocity ($\vmax$) of subhalos is hampered by the
fact that $\vmax$ for the observed Milky Way satellites is poorly
constrained by stellar velocity dispersion measurements (e.g.
\citealt{Strigari2007a}).

``Via Lactea'', the highest-resolution simulation of galactic
substructure to date \citep{Diemand2007a,Diemand2007b,Kuhlen2007}
offers the best opportunity for a systematic investigation of the
missing satellite problem. Several of the newly discovered ultra-faint
Milky Way satellites have estimated total masses just in excess of
$10^6\,\msun$ (like Leo IV, Coma Berenices, Canes Venatici II, see
\citealt{Simon2007}): such small dwarf galaxies are typically resolved
in Via Lactea with $>50$ dark matter particles, i.e. with a particle
mass that is two dex smaller than in \citet{Moore1999b}, three dex
smaller than in \citet{Klypin1999}, and more than one dex smaller than
in the \citet{Diemand2004} and \citet{Gao2004} simulations of galaxy
halos. Via Lactea follows the formation of a Milky Way-sized halo in
a {\it WMAP} 3-year cosmology: within its $\rtwo$ (the radius
enclosing an average density 200 times the mean matter value), more
than 2000 subhalos can be identified at $z=0$ with $\msub>4\times
10^6\,\msun$, distributed approximately with equal mass per decade of
mass.  Of these, 40 are found within 50 kpc, a region that appeared
practically smooth in previous simulations.  Interestingly, the number
of dwarf satellites in the inner halo of the Milky Way has recently tripled.
Six dwarfs are currently known within 50 kpc of the Galactic Center:
the LMC, Sagittarius, Ursa Major II, Coma Berenices, Willman 1, and
Segue 1 \citep{Mateo1998,Willman2006,Zucker2006, Belokurov2007}, for
an estimated total number (after correcting for the SDSS sky coverage)
of order 25. Note, however, that Willman 1 and Segue 1 lie at the
uncertain boundary between dark matter-dominated dwarfs and globular
clusters.

The apparent conflict between the Galaxy's relatively smooth stellar
halo and the extremely clumpy dark matter distribution of Via Lactea
raises a number of important questions. First and foremost, is the
amount of substructure found in Via Lactea typical of galaxy halos?
If so, how efficiently does gas accretion, cooling, and star formation
need to be suppressed as a function of subhalo mass to explain the
observed abundance of dwarfs? Because of tidal mass loss, many of the
subhalos that have small masses and circular velocities at the present
epoch were considerably more massive and could have formed stars in
the past.  Does a single gas removal/star formation quenching
mechanism determine the nature/fate of the subhalo population? What
subset of subhalos has properties similar to the known stellar
subsystems of the Milky Way? How does a dwarf's luminosity relate to
the mass evolutionary history of its host subhalo? Is the star
formation suppression mechanism a strong function of the environment,
or are small dark matter clumps inefficient at forming stars even
beyond the virial radius of the Milky Way?  We address some of these
issues here. The paper is organized as follows.  In \S\,2 we
compare the amount and distribution of Via Lactea substructure with
that measured in an elliptical-sized halo simulated with comparable
resolution in a different cosmology. In \S\,3 we show that, because of
late tidal mass loss, the number of subhalos that reached dwarf-galaxy
sizes before accretion onto Via Lactea is much larger than their
abundance today and the abundance of known Milky Way satellites.  If
substructure mass regulates star formation, then for a given mass
threshold many more subhalos should have been able to build a sizeable
stellar mass at some point in the past than indicated by their
present-day abundance.  Sections 4 and 5 discuss the main constraints
on mechanisms for the suppression of star formation in small subhalos.
The subhalo radial distribution and central dark matter densities
are discussed and compared to the observations in
\S\,6. Finally, we present a brief summary and our conclusions in
\S\,7.

\begin{figure*}
\epsscale{0.99}
\plotone{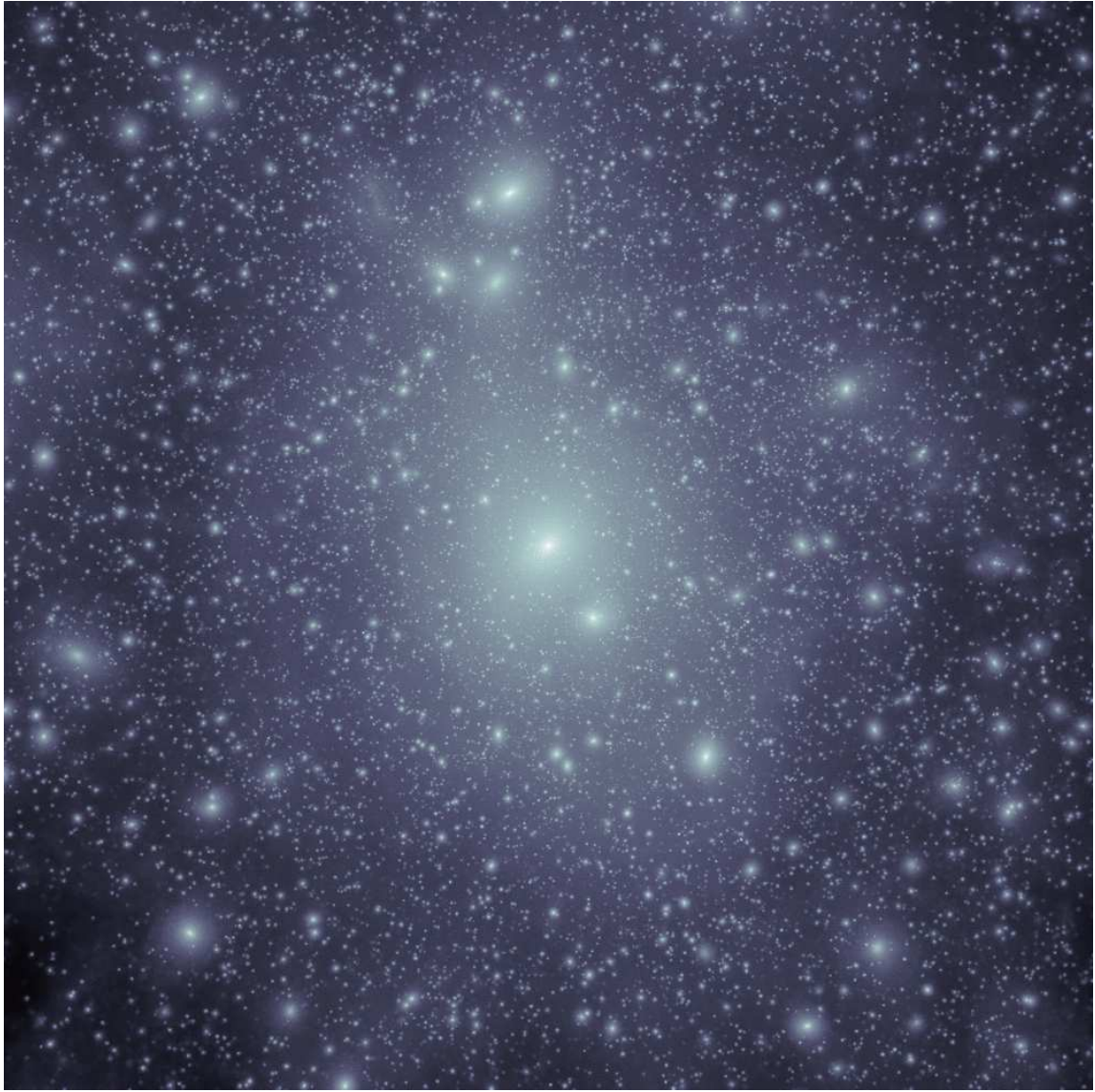}
\vspace{0.5cm}
\caption{Projected dark matter density-square map of our simulated
  elliptical-sized halo (``1e8Ell'') at $z=0.47$. The image covers an
  area of 980 $\times$ 980 physical kpc, and the projection goes
  through a 980 kpc-deep cuboid containing a total of 120 million
  particles and 25,000 identified subhalos. The logarithmic color
  scale covers 24 decades in density-square.  }
\label{elli}
\end{figure*}

\begin{figure*}
\plottwo{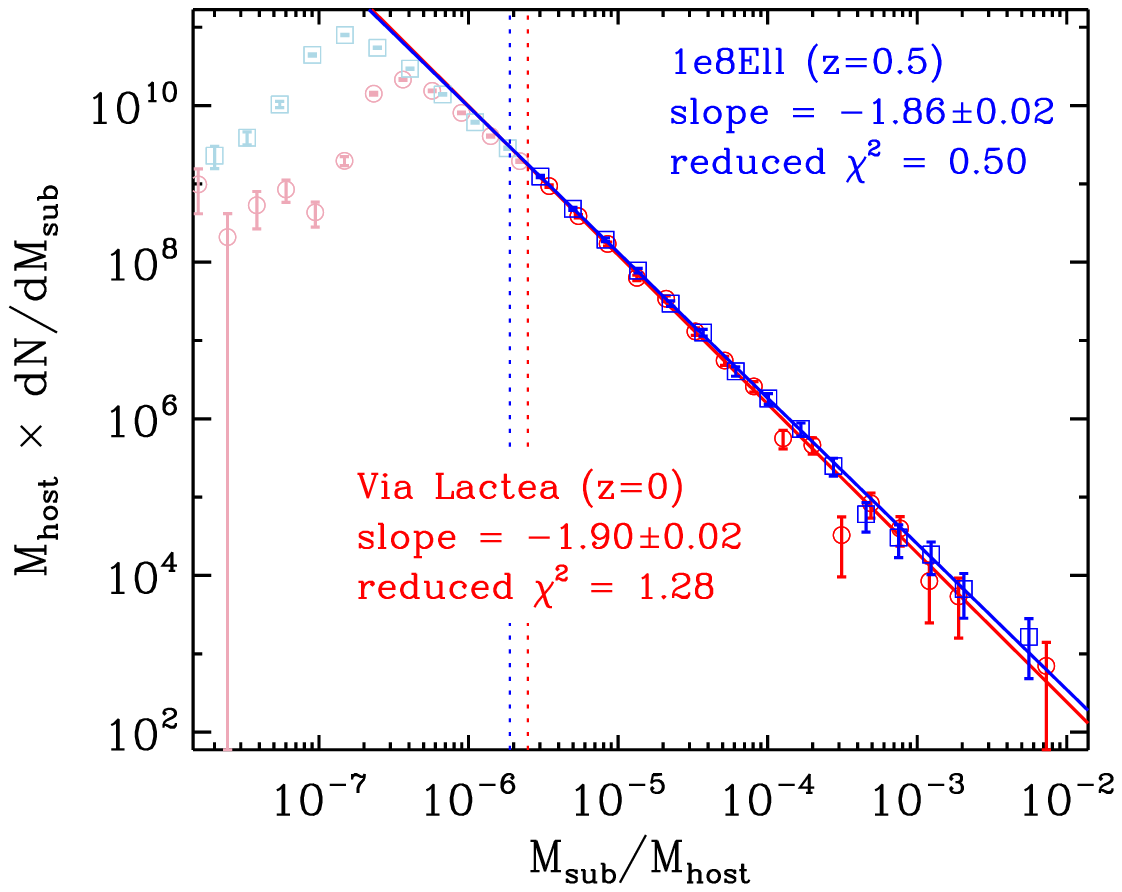}{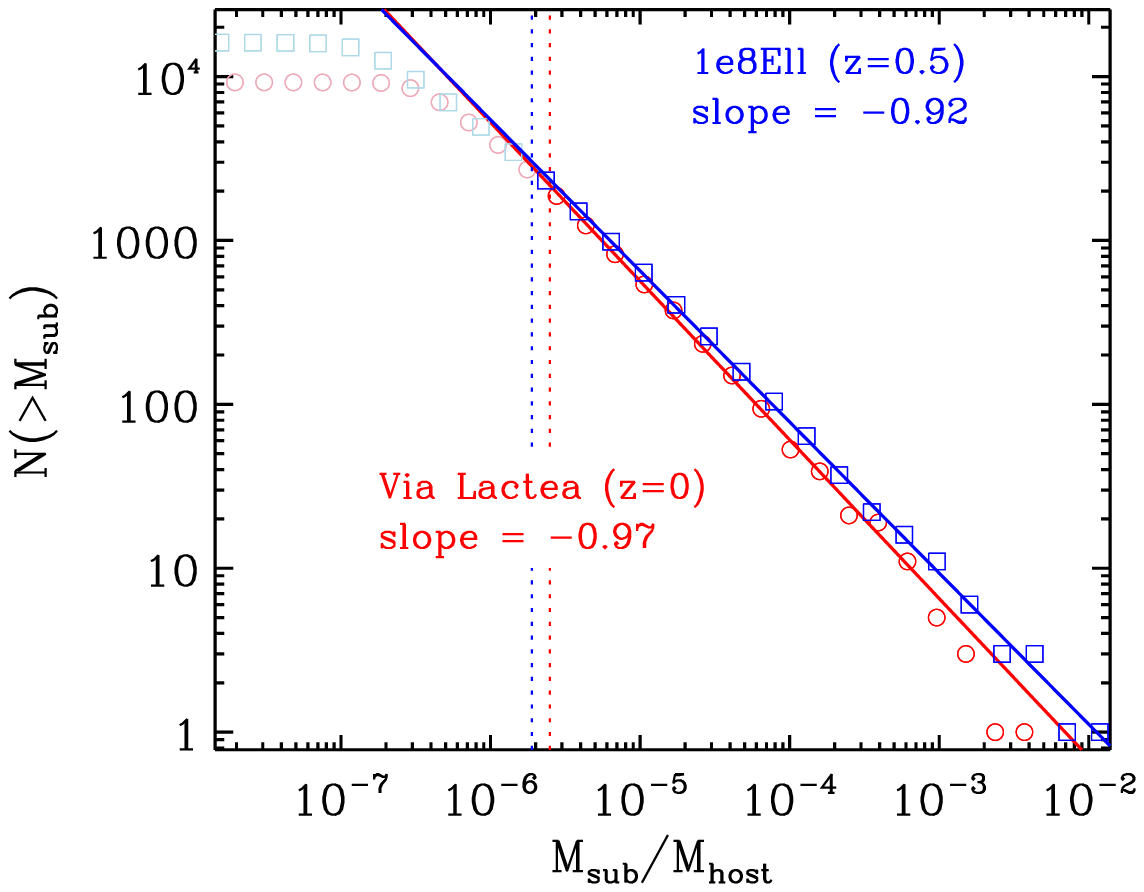}
\vspace{0.5cm}
\caption{Differential ({\it left panel}) and cumulative ({\it right
    panel}) subhalo mass functions within $\rtwo$ for 1e8Ell at
    $z=0.47$ ({\it blue square points}) and Via Lactea at $z=0$ ({\it
    red square points}), together with power-law fits ({\it solid
    lines}) in the range $200m_p<\msub<0.01\mhost$.  }
\label{massf}
\end{figure*}

\begin{figure*}
\plottwo{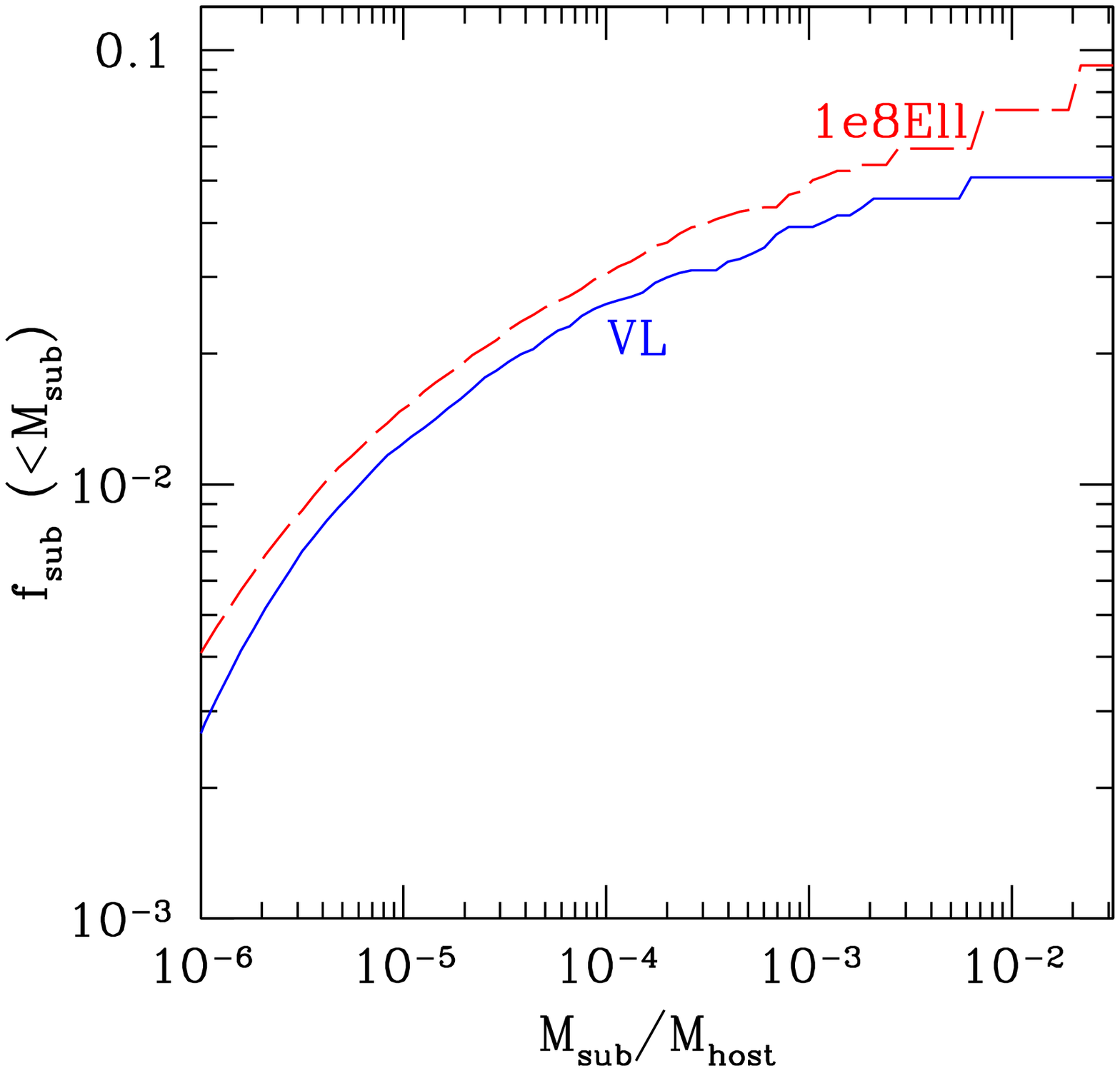}{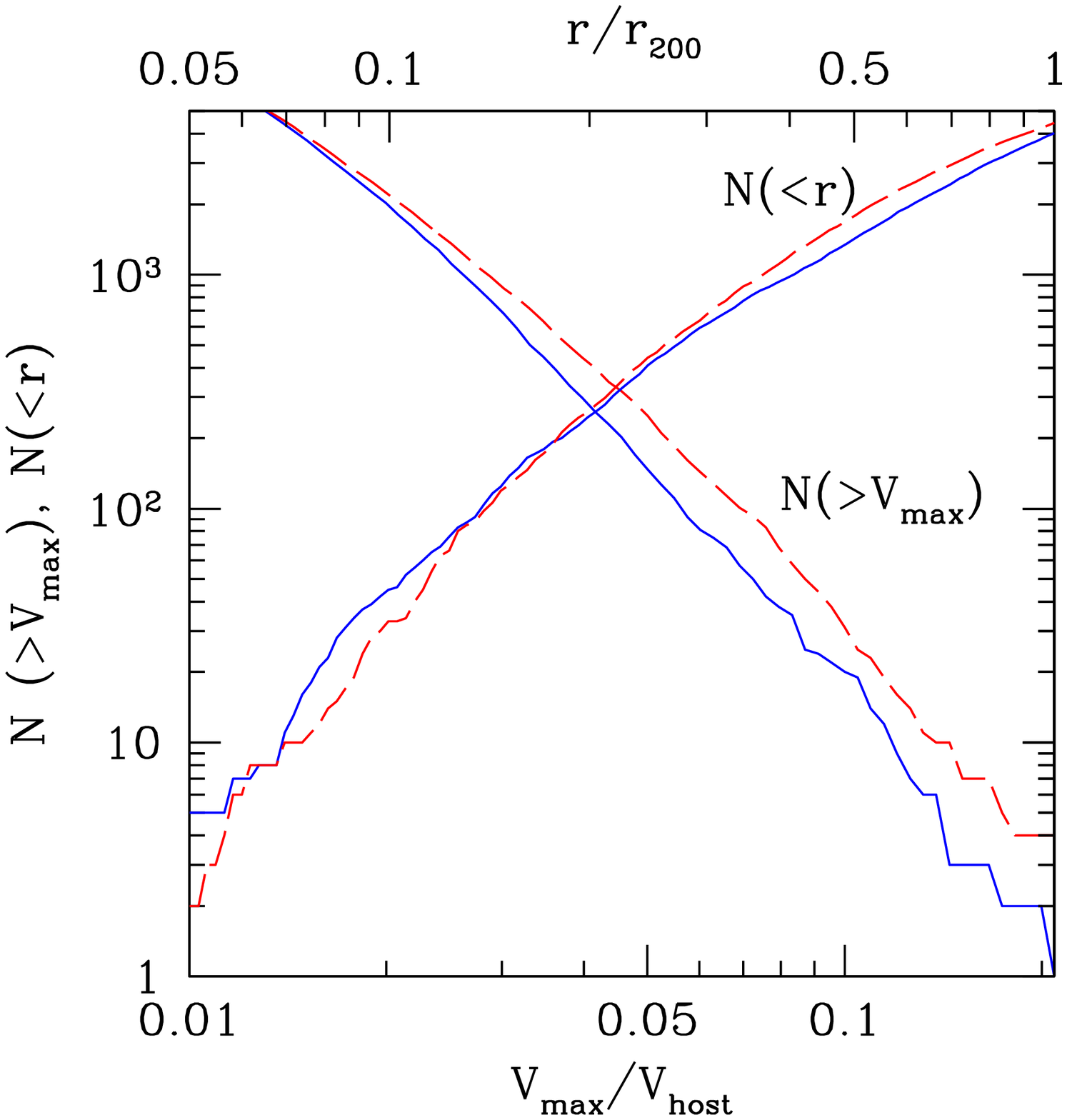}
\caption{{\it Left panel:} Fraction of the host halo mass that is
  bound up in substructure less massive than $\msub$ as a function of
  $\msub/\mhost$. {\it Lower solid curve}: Via Lactea. {\it Upper
    dashed curve}: 1e8Ell. {\it Right panel:} Cumulative maximum
  circular velocity function, $N(>\vmax)$, for all subhalos within
  $\rtwo$, as a function of $\vmax/\vhost$.  There are 7932 subhalos
  with $\vmax/\vhost>0.01$ in Via Lactea, versus 8005 in 1e8Ell.  Also
  plotted is the radial distribution, $N(<r)$, of all mass-selected
  subhalos with $\msub/\mhost>10^{-6}$ versus $r/\rtwo$. {\it Solid
    curves}: Via Lactea, $N(<\rtwo)=4021$. {\it Dashed curves}:
  1e8Ell, $N(<\rtwo)=4447$.  }
\label{fsub}
\end{figure*}

\section{CDM substructure in galaxy halos}

The Via Lactea simulation was performed with the PKDGRAV tree-code
\citep{Stadel2001} and employed multiple mass particle grid initial
conditions generated with the GRAFICS2 package
\citep{Bertschinger2001}. The high resolution region was sampled with
234 million particles of mass $m_p=2.1\times 10^4\msun$ and evolved
with a force resolution of 90 pc starting from redshift 50. We adopted
the best-fit cosmological parameters from the {\it WMAP} 3-year data
release \citep{Spergel2007}: $\Omega_M = 0.238$, $\Omega_\Lambda =
0.762$, $H_0= 73\,\kmsmpc$, $n=0.951$, and $\sigma_8=0.74$.  The $z=0$
host halo mass within $\rtwo=389$ kpc is $\mhost=1.77\times
10^{12}\,\msun$.  Its circular velocity profile reaches a peak of
$\vhost=181\,\kms$ at $\rvhost=62$ kpc.  Note that the size of the
Galaxy halo is still quite uncertain: since the effects of disk
formation on the total mass distribution are still poorly constrained,
the possible range of dark matter halo sizes that could form and host
a $220\,\kms$ disk is quite large, with $\vhost$ between 150 and
$220\,\kms$ \citep{Klypin2002}. When comparing Via Lactea to the Milky
Way, this uncertainty allows one to rescale velocities and distances
by $\pm 25\%$, and subhalo abundances above a given $\vmax$ by about a
factor of $1.25^3 = 2$. A comparable uncertainty in the abundances
comes from halo-to-halo scatter (e.g. \citealt{Reed2005}). Throughout
this paper we will for simplicity assume that Via Lactea matches the
properties of the Milky Way halo closely, but the substantial
uncertainties mentioned above should always be kept in mind.

For comparison purposes, we have run another simulation, termed
``1e8Ell'', that follows the formation of a more massive host halo
with $\mhost=1.34\times 10^{13}\,\msun$, $\rtwo=512\,$kpc, $\vhost=390
\kms$ and $\rvhost=116$ kpc.  The high resolution region of this second
run was sampled with 136,432,000 particles of mass $m_p=1.8\times
10^5\msun$ and evolved from redshift 59.5 down to $z=0.47$ with a
force resolution of 162 pc. For 1e8Ell the best-fit cosmological
parameters from the {\it WMAP} 1-year data release \citep{Spergel2003}
were adopted: $\Omega_M = 0.268$, $\Omega_\Lambda = 0.732$, $H_0=
71\,\kmsmpc$, $n=1$, and $\sigma_8=0.9$.

In both runs subhalos were identified using the phase-space
friends-of-friends algorithm 6DFOF described in \citet{Diemand2006}.
Around the centers of all groups with at least 16 members, spherical
density profiles were constructed, and the resulting circular velocity
profiles were fitted with the sum of a \citet*{NFW97} profile and a
constant density background. Subhalo tidal masses were assigned
according to the definition given in \citet{Diemand2007a}. The wealth
of substructure resolved in our elliptical-sized halo is comparable to
Via Lactea and is clearly seen in Figure \ref{elli}.  The halo finder
algorithm identifies comparable numbers of dark matter clumps within
the virial radius of Via Lactea at $z=0$ and that of 1e8Ell at
$z=0.47$.  Figure \ref{massf} shows the differential and cumulative
subhalo mass functions within $\rtwo$ for both runs. Substructure
extends down to the smallest mass scales reliably resolved: as shown
by \citet{Diemand2007a}, numerical resolution effects start to flatten
the mass distribution of subhalos below about 200 particles, and the
maximum circular velocity distribution below $\vmax/\vhost\approx
0.02$.  In the range $200m_p<\msub<0.01\mhost$, the best-fit slope of
the differential distribution, $dN/d\msub\propto \msub^{\alpha}$, is
$\alpha=-1.86\pm 0.02$ for 1e8Ell and $\alpha=-1.90\pm 0.02$ for Via
Lactea. In the same mass range the cumulative mass function has slope
$-0.92$ for 1e8Ell and $-0.97$ for Via Lactea.  Both simulations are
therefore characterized by steeply rising subhalo counts that, in the
case of Via Lactea, correspond to approximately equal mass per
substructure mass decade: most subhalos are of low mass. Figure
\ref{fsub} (left panel) depicts the fraction of the host halo mass
within a sphere of radius $\rtwo$ that is bound up in substructure
less massive than $\msub$, $f_{\rm sub}(<\msub)$ as a function of
$\msub/\mhost$. We measure a total mass fraction in substructure that
is about 5\% in Via Lactea and exceeds 9\% in 1e8Ell: its radial
distribution can be approximated as $f_{\rm sub}(<r)\propto r$ for
$0.1<r/\rtwo<1$.  Because of the steepness of the cumulative mass
function, these fractions appear to be converging rather slowly at the
small-mass end: more than 1\% of the host mass is found in clumps with
$\msub/\mhost<10^{-5}$. The amount of massive substructure is expected
to increase with host halo mass since more massive hosts form later
and dynamical friction and tidal-stripping have less time to operate
\citep{Gao2004,Zentner2005,vdB2005}. For the same reason, parent halos
of a given mass will have a larger abundance of subhalos at higher
redshifts than their present-day counterparts. However, when comparing
$z=0$ clusters with galaxy halos these effects are smaller than the
observed halo-to-halo scatter in subhalo abundance
\citep{Diemand2004,Reed2005}. Variation in the slope and normalization
of the power spectrum can also affect the amount of subhalos
significantly, especially their circular velocity function
\citep{Zentner2003}.

It is conventional to discuss the substructure population of galaxy
halos in terms of the circular velocity function. Figure \ref{fsub}
(right panel) shows the cumulative maximum circular velocity ($V_{\rm
  max} \equiv {\rm max} \{ \sqrt{G\msub(<r)/r} \}$) functions for the
entire subhalo population within the two hosts. These are well-fit by
a power-law of the form \be N(>\vmax)=100\,V_{\rm
  max,10}^{-3}~~~~~~(0.4<V_{\rm max,10}<3.5) \ee for Via Lactea, and
\be N(>\vmax)=1500\,V_{\rm max,10}^{-2.8}~~~~(0.8<V_{\rm max,10}<5.5)
\ee for 1e8Ell, with fractional errors not exceeding 20\% in both
cases.  (Here $V_{\rm max,10}\equiv \vmax/10$ $\kms$.) When
$\vmax$ is normalized to the maximum circular velocity of the host,
$\vhost$, the two substructure velocity functions are very similar at
the low end, and become more discrepant at larger $\vmax$ values.
There are 110 subhalos above $\vmax=10\,\kms=0.056\,\vhost$ in Via
Lactea, compared to 170 above $\vmax=22.5\,\kms=0.056\,\vhost$ in
1e8Ell.  As already discussed above, it is well known that large
subhalos (roughly within a factor of 30 of the maximum circular
velocity of the host) constitute a transient population that
continuously declines from mass-loss effects. The Via Lactea run has
now shown that small-scale substructure is instead more persistent and
independent of the age and size of the host \citep{Diemand2007b},
consistent with predictions from semi-analytic models
\citep{Zentner2005}. This is also supported by the left panel of
Figure \ref{fsub}: within the range $10^{-6}\lta \msub/\mhost \lta
10^{-2}$, the measured substructure mass fractions $f_{\rm
  sub}(<\msub)$ in the two simulations differ by less than $\sim
40$\%. While the similarity between the two objects is noteworthy, we
caution that such close agreement could be just a coincidence, and
substantial amount of halo-to-halo scatter may be present in the
abundance of small-scale substructure.\footnote{Note that the measured
abundance of small-scale substructure may also be sensitive to our
change from a {\it WMAP} year-1 (1e8Ell) to year-3 (Via Lactea)
cosmology, especially to the lower normalization of the
power-spectrum $\sigma_8$.}  The larger clumpiness of 1e8Ell is
dominated by its three surviving subhalos with $\msub/\mhost>0.006$,
versus none this massive identified in Via Lactea.

The right panel of Figure \ref{fsub} also shows the radial
distribution of subhalos as a function of $r/\rtwo$. In the range
$0.1<x\equiv r/\rtwo<1$, the abundance profiles of {\it mass-selected}
subhalos are very similar and well fit by \be N(<x)=N(<\rtwo){12
  x^3\over 1+11x^2}.  \ee This distribution agrees well with those
presented by \citet{Diemand2004} and \citet{Gao2004}, extending the
validity of these previous results down to $\msub>10^{-6}\,\mhost$.
Subhalos do not trace the matter distribution of the host: tidal
disruption is most effective in the inner halo, which leads to an
antibias in the abundance profile of substructure relative to the
smooth background, i.e. the radial distribution of subhalos above a
given mass is much flatter than that of the dark matter in the host
halo. This antibias is less pronounced if subhalos are selected by
today's peak circular velocity instead. The threshold
$\vmax/\vhost>0.02$ yields a sample of 2016 and 2334 subhalos in Via
Lactea and 1e8Ell, respectively. In the range $0.01<x<1$, the
abundance profiles of these {\it $\vmax$-selected} subhalos is well
fit by \be N(<x)=N(<\rtwo){20 x^{2.9}\over 1+19x^{1.9}}.  \ee

\section{Fossil records and missing satellites}

The Via Lactea and 1e8Ell runs discussed above show that, in the
standard CDM paradigm, galaxy halos are filled with tens of thousands
low-mass subhalos, about an order of magnitude more than found in
previous simulations. Subhalos lose mass from the outside in through
tidal stripping, the largest average mass fraction being lost at the
first pericenter passage \citep{Diemand2007b}. Larger clumps retain
less of their mass as they sink quickly towards the host center due to
dynamical friction, while the retained mass fraction is larger for
initially lighter subhalos that are less affected by dynamical
friction and stay in place. As many as 97\% of all well-resolved [$>5,000$
particles, $\vmax(z)>10\,\kms$] subhalos identified in Via Lactea at $z=1$ 
do not lose their identity and have a surviving $z=0$ remnant. This can be 
compared with the 47\% that survive from $z=4$ and the 20\% from $z=10$. 
Note that not all the subunits without a remnant today are totally 
disrupted to become part of the ``smooth" host halo, some simply merge into 
larger substructure that eventually survives down to the present epoch.   

\begin{figure*}
\plottwo{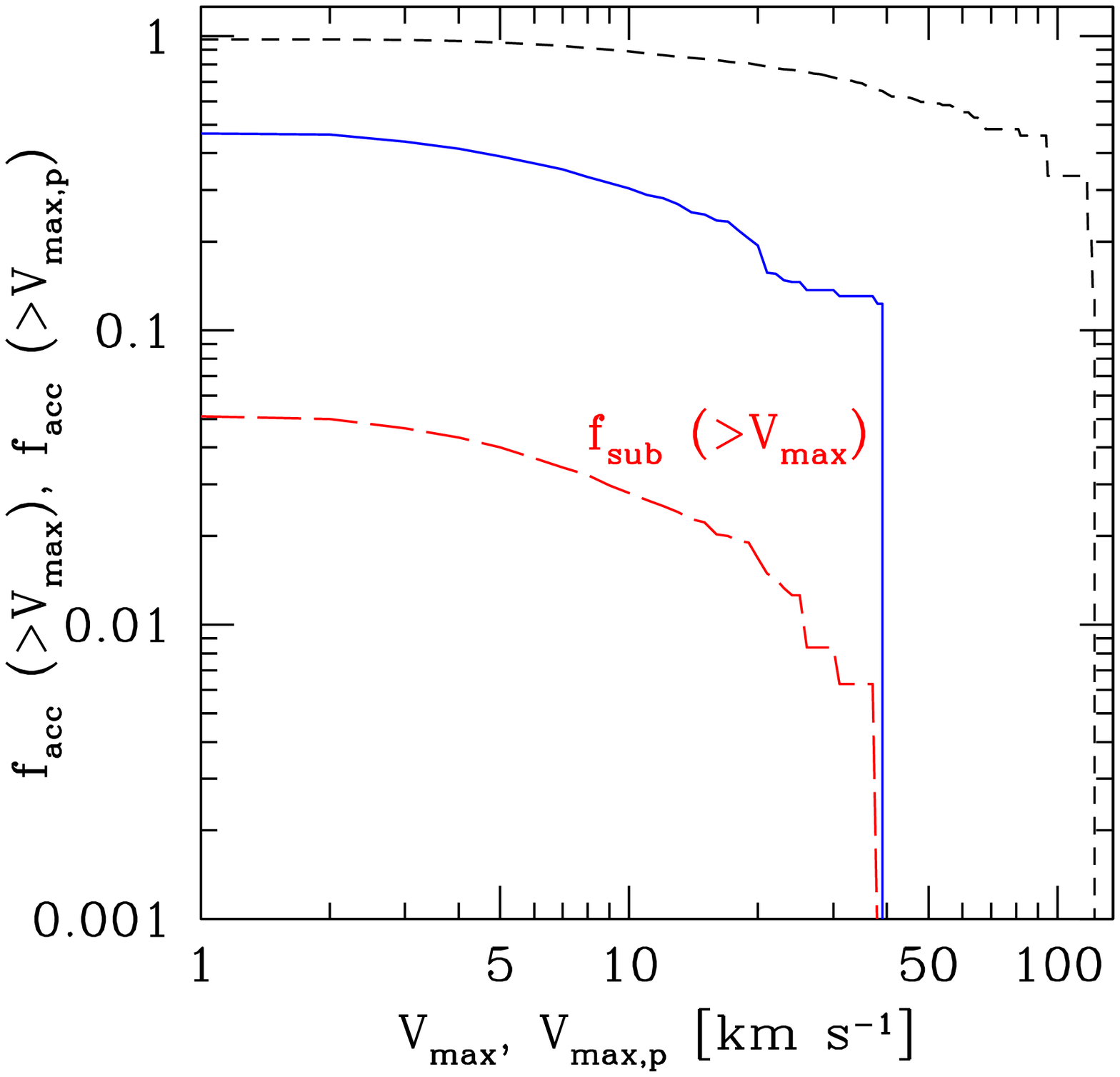}{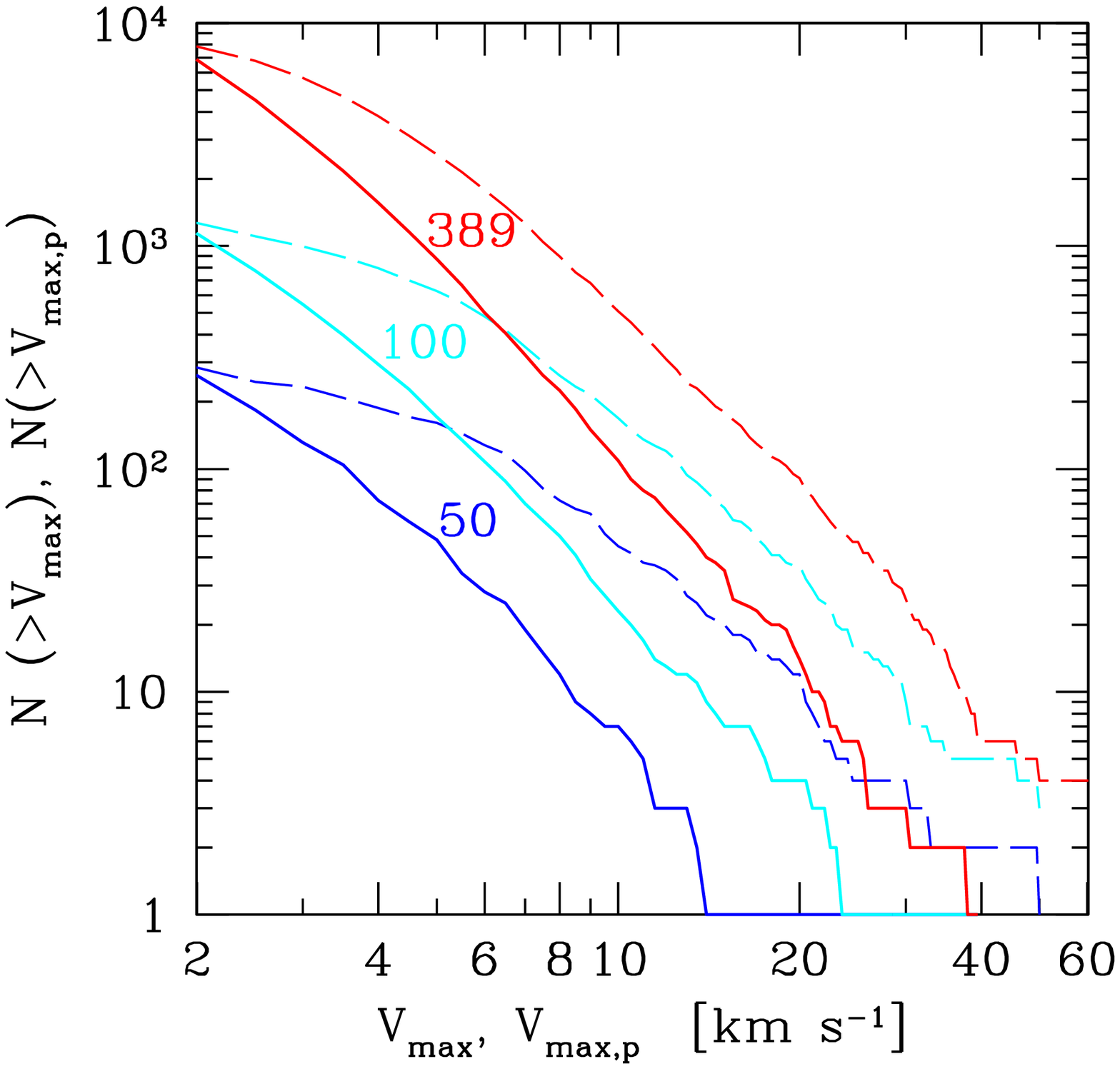}
\caption{{\it Left panel:} Via Lactea accreted mass fraction $f_{\rm acc}
 (>\vmax)$ (total mass brought in by all substructure {\it surviving today}
 within $\rtwo$ above $\vmax$, normalized to the mass of the host),
 as a function of today's subhalo circular velocity ({\it upper solid
 curve}). For comparison, we also plot the mass 
 $f_{\rm acc}(>\peakvmax)$ brought in by all resolved subunits  
 as a function of the highest maximum circular velocity reached
 throughout their lifetime, $\peakvmax$
 ({\it upper short-dashed curve}). The long-dashed lower curve
 shows the mass fraction $f_{\rm
 sub}(>\vmax)$ that remains in self-bound identifiable clumps at
 the present-epoch. {\it Right panel:}
 Cumulative circular maximum velocity function of all subhalos today
 within 50, 100, and 389 kpc ({\it solid lines, from bottom to top}).
 {\it Dashed lines:} Same for a sample of subhalos having a surviving
 bound remnant today and selected by their $\peakvmax$.
}
\label{fsub_peak}
\end{figure*}

It is interesting to know, at the high-resolution limit of our
simulations, what fraction of the Via Lactea's host halo mass was
accreted in identifiable clumps, and compare it with the mass 
brought in by subhalos that have a surviving bound remnant within
$\rtwo$ at the present epoch. The latter is expected to be significantly 
smaller than the former, since many major mergers at early times leave  
no surviving $z=0$ subhalos. We have therefore traced backward in time 
the progenitors of present-day remnants in each simulation snapshot, 
measured the {\it maximum mass} they had at any point during their 
history i.e. before they suffered tidal mass losses, summed up 
this mass over all subhalos within $\rtwo$ having peak circular 
velocity $>\vmax$, and divided the result by the mass of Via Lactea 
today. The total mass fraction, $f_{\rm acc}(>\vmax)$, brought in 
by surviving substructure is plotted in Figure \ref{fsub_peak} as a 
function of $\vmax$, and is found to exceed 45\% for $\vmax>2\,\kms$. 
There are $(14,109,873)$ subhalos today with $\vmax>(20,10,5)\,\kms$, which
contributed $(19,30,39)$\% of the mass of Via Lactea. Only 9-10\% of
the total material they brought in remains in self-bound identifiable clumps
at the present-epoch, the rest having been removed by tidal effects to
become part of the ``smooth'' halo component. 
Note that numerical effects lead to less cuspy subhalos that suffer from
artificially high-mass loss, especially in the inner halo where tidal
forces are strong \citep{Kazantzidis2004}. Figure \ref{fsub_peak} also 
shows the total mass accreted in clumps, $f_{\rm acc}(>\peakvmax)$, 
independently of whether these survive or not to the present day. 
This is plotted against the highest $\vmax$ reached by subunits
throughout their lifetime, a quantity we refer to as 
`$\peakvmax$' (see also \citealt{Kuhlen2007}). Interestingly, more than 97\% 
of the Via Lactea mass is acquired in resolved discrete clumps with 
no evidence for significant smooth accretion. 


\begin{figure*}
\epsscale{0.85}
\plotone{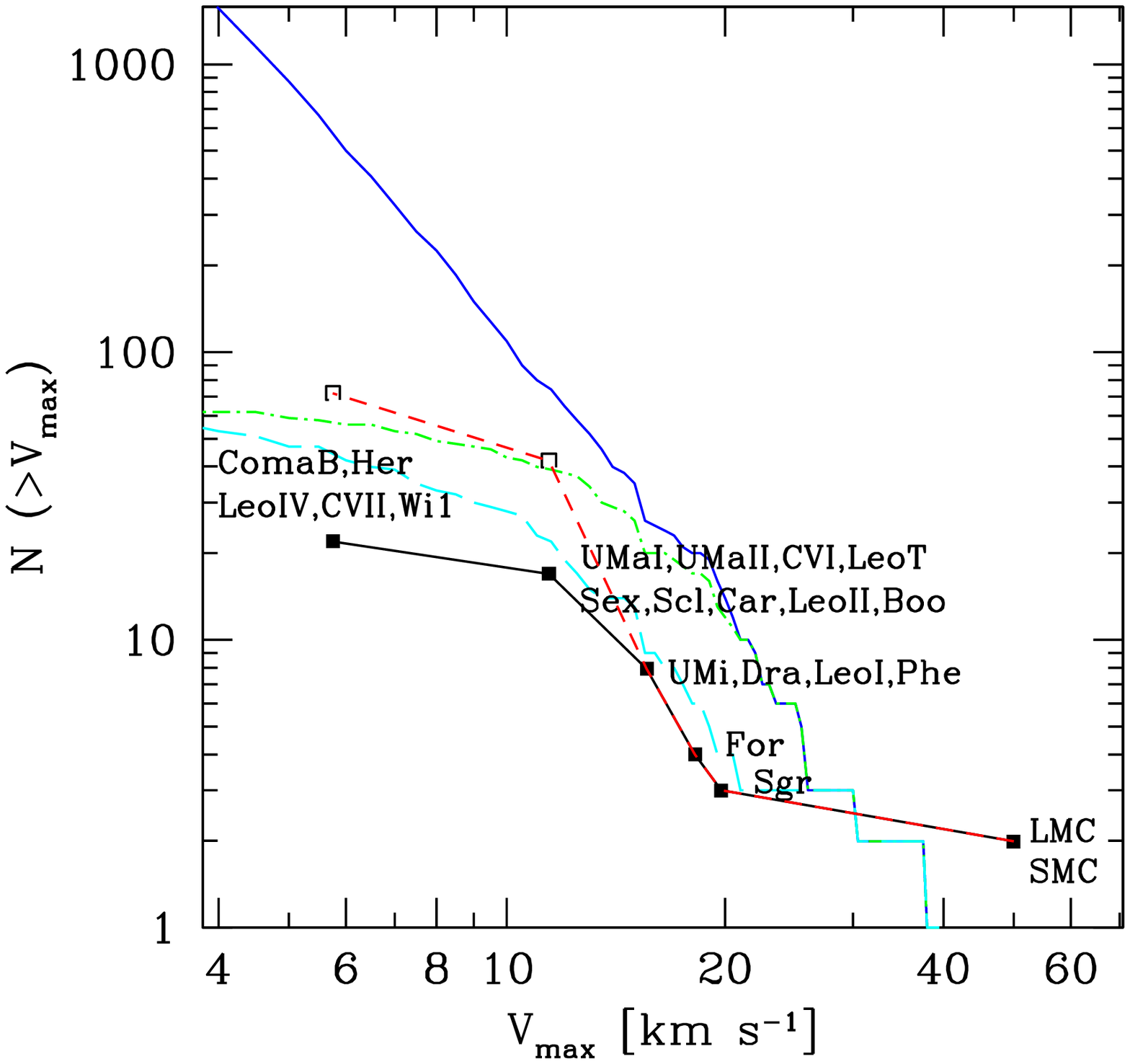}
\caption{Cumulative number of Via Lactea subhalos within $\rtwo$ ({\it
    solid curve}) as well as all Milky Way satellite galaxies within
  420 kpc ({\it filled squares}), as a function of circular velocity.
  The data points are from \citet{Mateo1998}, \citet{Simon2007},
  \citet{Munoz2006}, and \citet{Martin2007}, and assume a maximum
  circular velocity of $\vmax=\sqrt{3}\sigma$ \citep{Klypin1999}. The
  short-dashed curve connecting the empty squares shows the expected
  abundance of luminous satellites after correcting for the sky
  coverage of the SDSS. {\it Dash-dotted curve:} circular velocity
  distribution for the 65 largest $\peakvmax$ subhalos before
  accretion (LBA sample). {\it Long-dashed curve:} circular velocity 
  distribution for the ``fossil of
  reionization'' EF sample. This includes the 61 largest (sub)halos at
  $z=13.6$ [$\vmax(z=13.6)>4\,\kms$] plus the 4 (sub)halos that reach
  a $\peakvmax>38\,\kms$ after the epoch of reionization and are not
  in the largest 61 at $z=13.6$.
}
\label{velf_sat}
\end{figure*}

The large tidal mass losses suffered by massive subhalos have obvious
implications for astrophysical solutions to the satellite problem. The
leading scenario for reconciling the mismatch between luminous and
dark substructure in the Milky Way posits that stars are able to form
continuously and efficiently only in halos above a certain mass
threshold, a threshold determined by the inability of shallow
potential wells to accrete intergalactic gas photoionized by the
ultraviolet (UV) background and/or to retain gas blown out by
supernova explosions. There are slightly in excess of 20 dwarf
galaxies known to orbit the Milky Way. After correcting for the sky
coverage of the fifth data release of the SDSS, i.e. after weighing
each of the new ultra-faint dwarfs by a factor $\sim 6$, the total
number of satellites is probably around 60-70
\citep{Simon2007,Koposov2007} down to a surface brightness limit of
$\mu_V=30$ mag arcsec$^{-2}$. A solution to the substructure problem
suggested by \citet{Stoehr2002} and \citet{Hayashi2003} would be to
place the luminous dwarfs in the most massive subhalos {\it at the
  present epoch}, suitably adjusting the mass (or peak circular
velocity) threshold for efficient galaxy formation.  There are 65
subhalos today in Via Lactea (within $\rtwo$) with $\msub>1.4\times
10^8\,\msun$ or $\vmax>12\,\kms$: the paucity of Milky Way satellites
would then be simply explained by the inhibition of star formation
below the above values of $\msub$ and $\vmax$. Baryonic material
falling in the dark matter potential well will be shock heated to the
effective virial temperature of the host, compressed, and eventually
cool and fragment into stars. A circular velocity threshold of
$12\,\kms$ corresponds to a virial temperature \be T_{\rm vir}={\mu
  m_p \vmax^2\over 2k_B}<8700\,\mu\,{\rm K},
\label{tvir1}
\ee where $\mu$ is the mean molecular weight. Figure \ref{velf_sat}
shows the cumulative number of Via Lactea subhalos as well as all
Milky Way satellite galaxies within 420 kpc (the distance of Leo T,
\citealt{Irwin2007}) as a function of circular velocity. The current
available data are summarized in Table 1. The data points in the
figure include all the previously known dwarfs \citep{Mateo1998} plus
the new circular velocity estimates of the ultra-faint Milky Way
satellites from \citet{Simon2007}, plus Bo\"{o}tes \citep{Munoz2006}
and Willman 1 \citep{Martin2007}. They have been plotted assuming a
maximum circular velocity of $\vmax=\sqrt{3}\sigma$
\citep{Klypin1999}, where $\sigma$ is the measured stellar
line-of-sight velocity dispersion, i.e.  assuming a stellar spherical
density profile $\propto r^{-3}$ in a singular isothermal potential.
Note that the assumption of a constant multiplicative factor between
$\vmax$ and $\sigma$ is merely the simplest thing to do, and is not
likely to hold on a case-by-case basis. Detailed modeling of the
radial velocity dispersion profile, allowing for variations in the DM
mass distribution and the stellar velocity anisotropy
\citep{Strigari2007a}, would be preferrable, but is currently only
available for a subset of all known dwarfs.

\begin{deluxetable*}{l c c c c l}
\tablewidth{0pt}
\tablecolumns{6}
\tablecaption{The Milky Way satellites}
\tablehead{
\colhead{Name} & \colhead{Heliocentric distance} & \colhead{$\sigma$} 
& \colhead{$\rho_0$} & \colhead{Core radius $r_c$} & \colhead{References} \\
\colhead{} & \colhead{(kpc)} & \colhead{($\kms$)} & \colhead{($\sden$)} &
\colhead{(pc)} & \colhead{}}

\startdata

 Segue 1           & 23 & -- & -- & 19 & (1)\\
 Sagittarius       & 24 & 11.4 & 0.07 & 550& (2)\\
 Ursa Major II     & 32 &  6.7 & 1.14 & 81 & (3),(4)\\ 
 Willman 1         & 38 & 4.3 & 18 & 13 & (5),(6)\\ 
 Coma Berenices    & 44 & 4.6 & 2.1 & 41 & (1),(3)\\ 
 LMC               & 49 & 20.2 & -- & -- & (2),(7)\\ 
 SMC               & 58 & 27.5 & -- & -- & (2),(8)\\ 
 Bo\"{o}tes        & 60 & 6.6 & 0.34 & 145 & (9),(10) \\
 Bo\"{o}tes II     & 60 & -- & -- & 46 & (11) \\
 Ursa Minor        & 66 & 9.3 & 0.36 & 200 & (2)\\ 
 Sculptor          & 79 & 6.6 & 0.60 & 110 & (2)\\
 Draco             & 82 & 9.5 & 0.46 & 180 & (2)\\ 
 Sextans           & 86 & 6.6 & 0.06 & 335 & (2)\\
 Carina            & 101& 6.8 & 0.17 & 210& (2) \\ 
 Ursa Major I      & 106& 7.6 & 0.25 & 197& (3)\\
 Fornax            & 138& 10.5 & 0.09 & 460 & (2)\\
 Hercules          & 138& 5.1 & 0.10 & 205 & (1),(3)\\ 
 Canes Venatici II & 151& 4.6 & 0.49 & 85 & (1),(3) \\
 Leo IV  & 158 & 3.3 & 0.19 & 97 & (1),(3)\\         
 Canes Venatici I  & 224 & 7.6 & 0.08 & 355 & (3)\\ 
 Leo II  & 233 & 6.7 & 0.24 & 178 & (2),(12),(15)\\
 Leo I & 255 & 9.2 & 0.23 & 245 & (13),(16)\\
 Leo T & 417 & 7.5 & 0.79 & 109& (3),(17)\\             
 Phoenix &420 & 8.9 & 0.14 & 310& (2),(14)\\ 
\enddata
\tablerefs{(1) \citealt{Belokurov2007}; (2) \citealt{Mateo1998}; (3) \citealt{Simon2007}; 
(4) \citealt{Zucker2006}; (5) \citealt{Willman2006}; (6) \citealt{Martin2007}; 
(7) \citealt{vdM2002}; (8) \citealt{Harris2006}; (9) \citealt{Belokurov2006}; 
(10) \citealt{Munoz2006}; (11) \citealt{Walsh2007}; (12) \citealt{Bellazzini2005};
(13) \citealt{Mateo2007}; (14) \citealt{Young2007}; (15) \citealt{Coleman2007}; 
(16) \citealt{Bellazzini2005}; (17) \citealt{Irwin2007}.
}
\label{targettable}
\end{deluxetable*}

If the stellar systems deeply embedded in dwarf spheroidals remain
largely unaffected by tidal stripping (this is clearly not the case,
e.g., for Ursa Major II and Sagittarius), then the mass removal of
large fractions of their original halo mass by tidal effects may make
solutions in which luminosity tracks {\it current} subhalo mass
somewhat misleading. Our simulations show and quantify better than
before that many of the dark matter clumps that have small masses and
circular velocities at the present epoch were considerably more
massive and should have formed stars in the past (e.g.
\citealt{Kravtsov2004}). We illustrate this point in Figure
\ref{fsub_peak} (right panel), which shows the cumulative circular
maximum velocity function of substructure within 50, 100, and 389 kpc.
Also plotted, for comparison, is the abundance of {\it surviving}
subhalos selected instead by the highest circular velocity they reached
throughout their lifetime, $\peakvmax$. Subhalos will reach 
their $\peakvmax$
at a redshift $z_{\rm max}$ before falling into Via Lactea: this type
of circular velocity selection is designed then to remove the bias
introduced by tidal mass losses, and {\it to highlight the subhalos
  that may have started shining before being accreted by their host.}
Within $\rtwo$, the number of massive galactic subhalos that reached a
peak circular velocity in excess of $10\,\kms$ at some point during
their history is 510, about {\it five times larger} than their
present-day abundance. This ratio increases with increasing
$\peakvmax$ and decreasing radius: a) above a virial temperature
$T_{\rm vir}=10,000$ K, or a circular velocity $\vmax=16.7\,\kms$
($\mu=0.59$ for fully ionized primordial gas), gas can cool
efficiently and fragment via excitation of hydrogen Ly$\alpha$. The
number of subhalos within $\rtwo$ that reached this ``atomic cooling''
mass at some point in the past is 135, nearly {\it six times larger}
than their present-day abundance; b) within the inner 50 kpc there is
today only one subhalo with $\vmax>16.7\,\kms$, but there are 16
surviving remnants that had this peak circular velocity and were more
massive at earlier times.  {\it If substructure mass regulates star
  formation, then for a given mass threshold many more subhalos should
  have been able to build a sizeable stellar mass at some point in the
  past than indicated by their present-day abundance.}

It is important then to investigate the consequences of a mass (or
circular velocity) cut that picked instead the top (say) 65 most
massive (or largest $\peakvmax$) subhalos at all epochs as the hosts
of the known Milky Way dwarfs. Following \citet{Diemand2007b}, such or
similar samples have been termed ``LBA'' (for ``largest before
accretion''subhalos) by \citet{Strigari2007a} and \citet{Simon2007}.
The idea behind this selection is to allow star formation only above a
relatively large constant critical size, a scenario of permanently
inefficient galaxy formation in all smallest systems, independently of
time-varying changes in the environment like those triggered, e.g., by
reionization. Today's circular velocity distribution of our LBA sample
is shown in Figure \ref{velf_sat}: interestingly, this sample includes
12 of the 14 subhalos above $\vmax=20\,\kms$ identified today, and 26
of the 35 identified above $\vmax=15\,\kms$, i.e.  the most massive
today and LBA subpopulations basically coincide at large values of
$\vmax$.\footnote{Note that the same is not true for the top 10 LBA
  subhalos \citep{Kravtsov2004,Diemand2007b,Strigari2007a}, as the
  largest $\peakvmax$ systems suffer the largest mass loss and are
  removed from the top ten list of more massive systems at $z=0$.}\
\textit{Therefore a solution to the substructure problem in which only
  the largest 50-100 $\peakvmax$ subhalos at all epochs were able to
  form stars efficiently would automatically place the luminous Milky
  Ways dwarfs in the most massive subhalos at the present epoch.}  To
match the circular velocity function of the LBA sample, however, the
observed dwarf spheroidals (dSphs) must have circular velocity
profiles that peak at values well in excess of the stellar velocity
dispersion (see Fig. \ref{velf_sat} and discussion below).  Note that
the cut in $\peakvmax$ instead of $\vmax$ of the LBA sample requires
star formation to be inhibited in all subhalos with
$\peakvmax<21.9\,\kms$ or virial temperature \be T_{\rm vir}={\mu m_p
  \peakvmax^2\over 2k_B}<17,000\,{\rm K}.
\label{tvir2}
\ee

\section{Suppressing dwarf galaxy formation}

The two thresholds for efficient star formation given in equations
(\ref{tvir1}) and ({\ref{tvir2}) provide the correct total number of
  luminous Milky Way satellites (assumed to be around 60-70), not a
  match to the observed circular velocity function. A careful look at
  Figure \ref{velf_sat} suggests two possible solutions to the
  mismatch problem:

\begin{enumerate}

\item stars in the Milky Way dSphs are deeply embedded within their
  dark matter halos.  The halo circular velocity profiles peak well
  beyond the luminous radius at speeds significantly higher that
  expected from the stellar line-of-sight velocity dispersion, i.e.
  $\vmax\sim 3\sigma$ as suggested by \citet{Stoehr2002} and
  \citet{Penarrubia2007}.  This scenario would shift the data points
  in Figure \ref{velf_sat} by about a factor $\sqrt{3}$ further to the
  right, making the mass distribution of the luminous Milky Way dwarf
  spheroidals agree with the steep mass function of the most massive
  Via Lactea subhalos today. Its main drawback is that it boosts the
  number of massive {\it inner} satellites to values that are
  difficult to reconcile with $\Lambda$CDM halos.  There are five
  Milky Way dwarfs within 50 kpc of the Galactic Center with measured
  stellar velocity dispersions $\sigma$ above $4\,\kms$: the SDSS
  ultra-faint Coma Berenices, Willman 1, and Ursa Major II, together
  with the LMC and Sagittarius. Correcting for the sky coverage of the
  SDSS and using the conversion $\vmax\sim 3\sigma$ would bring the
  total number of inner satellites above $\vmax\sim 12\,\kms$ to
  twenty or so. By contrast, there are only three subhalos in Via
  Lactea with $d<50$ kpc and $\vmax>12\,\kms$ today (see the right
  panel of Fig. \ref{fsub_peak}). Note that, even at the resolution of  
  Via Lactea, subhalos within (say) 50 kpc from the center may still
  be subject to numerical overmerging. We will compare in detail the radial
  distribution of Milky Way's satellites and Via Lactea subhalos in
  the next section.

\item if $\vmax\sim \sqrt{3}\sigma$ instead, then the discrepancy
  between the number of {\it massive} substructure expected in
  $\Lambda$CDM and the known companions of the Milky Way can be solved
  if only approximately one out of five subhalos with $\vmax>20\,\kms$
  {\it today} houses a luminous dwarf satellite. Because of dynamical
  friction and tidal mass losses, the large subhalos that are today
  devoid of stars were typically more massive in the past. This
  scenario therefore requires star formation to be quenched in objects
  that reached peak circular velocities as large as $35-40\,\kms$ in
  the past.

\begin{figure*}
\plottwo{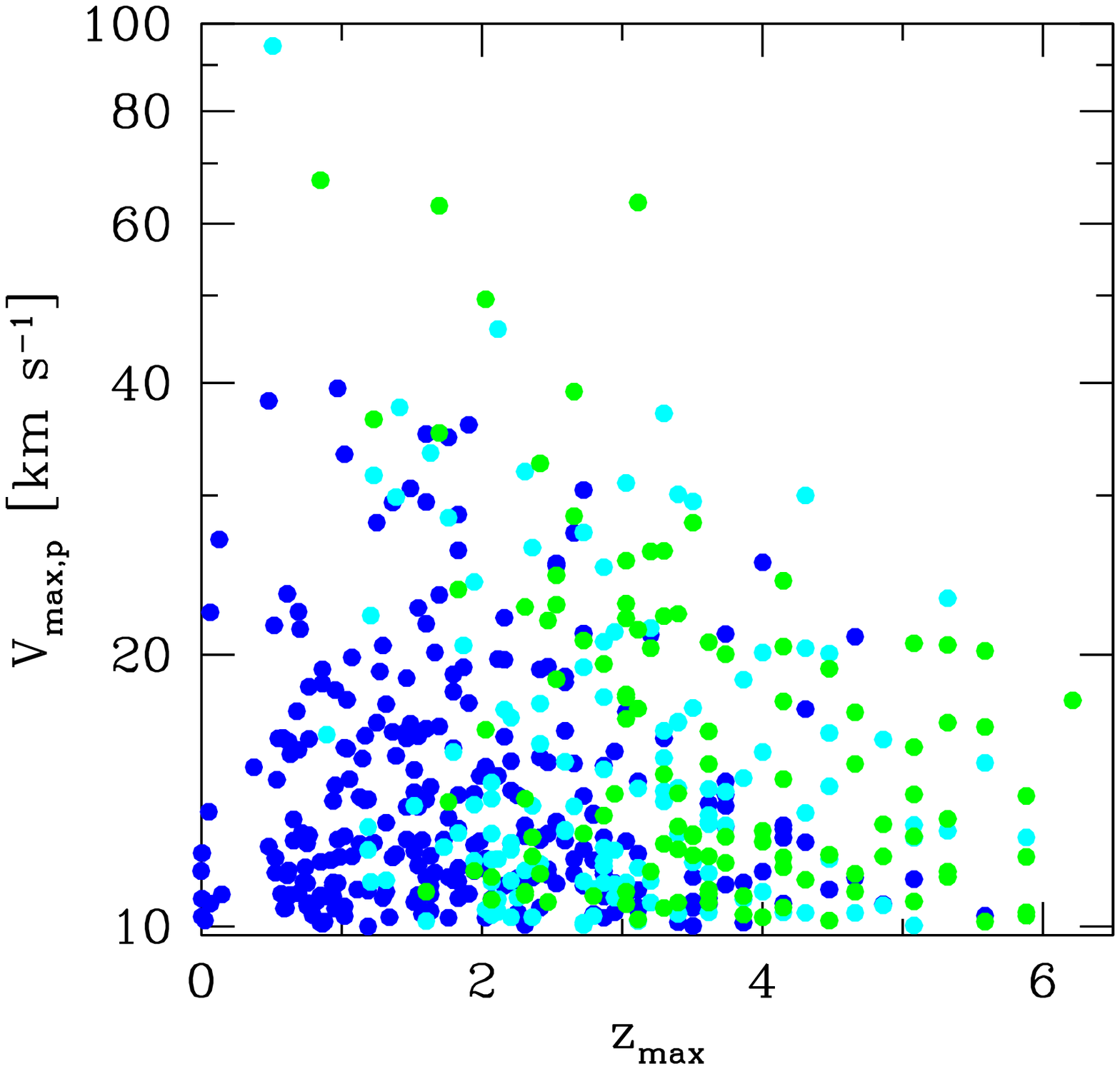}{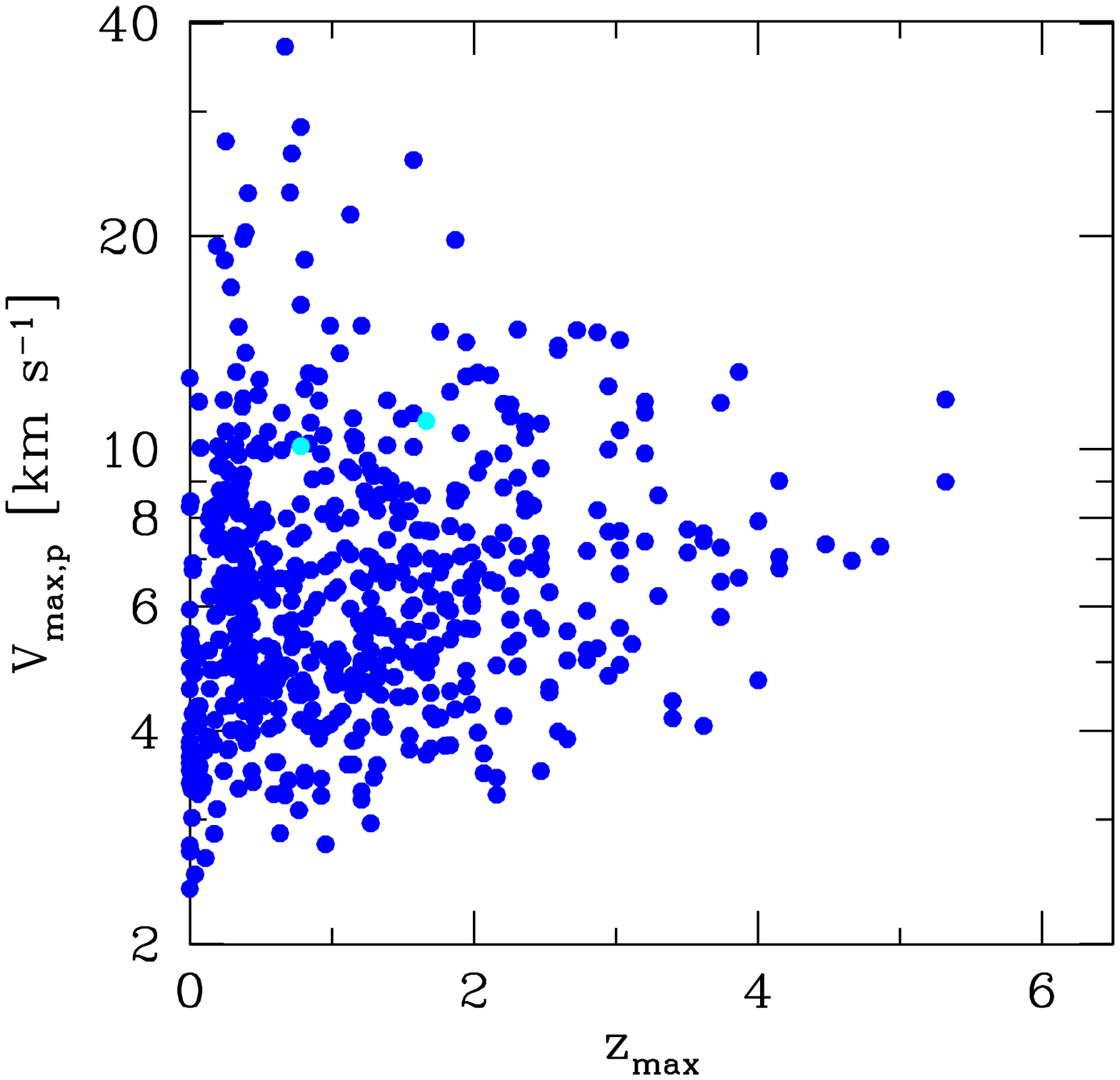}
\caption{{\it Left panel:} All (510) Via Lactea subhalos that reached
  a $\peakvmax>10\,\kms$ at redshift $z_{\rm max}$, plotted in the
  $\peakvmax-z_{\rm max}$ plane. The color coding indicates the
  different intervals of $\vmax$ where those subhalos end up today:
  $\vmax>\peakvmax/2$ ({\it blue dots}),
  $\peakvmax/3<\vmax<\peakvmax/2$ ({\it cyan dots}),
  $\vmax<\peakvmax/3$ ({\it green dots}). Clumps that were accreted
  earlier ($z_{\rm max}>3$) are more heavily affected by tidal
  stripping and their $\peakvmax$'s drop by more than 67\%.  {\it
    Right panel:} Same (but on a different $\peakvmax$ scale) for the
  sample of 575 halos that lie today in the ``field'' between 389 kpc
  ($\rtwo$) and 584 kpc ($1.5\rtwo$) and are more massive than
  $10^7\,\msun$. Most of these are accreted late ($z_{\rm max}\lta 1$)
  and suffer only moderate mass loss.  }
\label{vpeak_z}
\end{figure*}

\end{enumerate}

To shed further light on the star formation suppression mechanism that
is acting on galactic substructure we have plotted in Figure
\ref{vpeak_z} (left panel) all Via Lactea subhalos that reached a
$\peakvmax>10\,\kms$ at redshift $z_{\rm max}$. Out of a total of 510
subhalos, 318 reached their $\peakvmax$ at redshift $z_{\rm max}>2$,
180 at $z_{\rm max}>3$, and 80 at $z>4$. The color coding indicates
the different intervals of $\vmax$ where those subhalos end up today.
Clumps that were accreted earlier ($z_{\rm max}>3$) are more heavily
affected by tidal stripping and their $\peakvmax$'s drop by more than
67\%.  If stars can form continuously and efficiently only above a
mass or circular velocity threshold, then clearly \textit{any
  quenching mechanism must already be in place at early times}.

At a distance of 420 kpc and with a surface brightness of $\mu_V=26.9$
mag arcsec$^{-2}$, Leo T has well-measured kinematics from both the
stars and the gas, a stellar velocity dispersion of $7.5\pm
1.6\,\kms$, an estimated mass just short of $10^7\,\msun$, an absolute
magnitude $M_V=-7.1$, and a total mass-to-light ratio ($M/L_V$) of
$138\pm 71$ \citep{Irwin2007,Simon2007,Ryan2007}. The detection
efficiency for satellites this luminous over the SDSS survey area is
nearly unity, so only a handful of similar objects is expected over
the whole sky. Even at distances well in excess of 500 kpc, objects
brighter than $M_V=-8$ would have been found by the SDSS
\citep{Koposov2007}.  In the right panel of Figure \ref{vpeak_z} we
have plotted in the same $\peakvmax-z_{\rm max}$ plane (but on a
different $\peakvmax$ scale) the sample of 575 halos that lie today in
the ``field'' between 389 kpc ($\rtwo$) and 584 kpc ($1.5\rtwo$) and
are more massive than $10^7\,\msun$. And while this numerous
subpopulation may be shining just under the detection threshold as
ultra-faint dwarfs, it seems fair to conclude that {\it small dark
  matter clumps must be relatively inefficient at forming stars even
  well beyond the virial radius}. The figure shows that most of these
``field'' halos are newcomers, i.e. they are accreted late ($z_{\rm
  max}\lta 1$) and suffer only moderate mass loss. Contrary to what is
invoked for more massive dwarf spheroidals sinking within 50-100 kpc
from the Galactic Center \citep{Mayer2007}, gas stripping by ram
pressure in a hot low-density Galactic corona is unlikley to be
responsible for the truncation of star formation in this
subpopulation. Indeed, while many of these field halos are former
subhalos, i.e. passed through the host at some earlier time, only 15\%
of them had pericenters within 50 kpc.

\section{A toy model of reionization suppression}

A popular model for the suppression of gas accretion and star
formation in small halos is the ``reionization scenario''. In this
hypothesis the observed dSphs correspond to the surviving substructure
population that accreted a significant amount of gas before the epoch
of reionization at redshift $>10$ and achieved a potential well deep
enough to form stars: after reionization, UV photoionization feedback
results in the quenching of star formation in all newly forming
subhalos below some mass threshold. While this general model has been
investigated by a number of authors in the recent literature (e.g.
\citealt{Bullock2000,Benson2002,Somerville2002,Kravtsov2004,Ricotti2005,Moore2006,
  Diemand2007a,Strigari2007a,Simon2007}), there are considerable
differences in the specifics of its many variants.

The {\it Wilkinson Microwave Anisotropy Probe (WMAP)} 3-year data
require the universe to be fully reionized by redshift $z=11\pm 2.5$
(Spergel \etal 2006). The reionization history of a given 
galaxy-forming region appears to depend strongly on the mass of its
present-day halo, with higher mass systems forming earlier on average
and thus being reionized earlier \citep{Weinmann2007}. Following 
\citet{Moore2006}, let us assume that
the region around the Milky Way was reionized around $z=12-13$, i.e.
when reionization was well advanced in the universe as a whole.
Cosmological hydrodynamic simulations of early structure formation in a
$\Lambda$CDM universe have found that a typical H$_2$ fraction in
excess of 200 times the primordial value is produced after the
collapse of halos with virial temperatures above 1000 K.  This is
large enough to efficiently cool the gas and allow it to collapse and
form stars within a Hubble time unless significant heating occurs
during this phase (e.g.
\citealt{Machacek2001,Yoshida2003,Kuhlen2005}). A virial temperature
of 1200 K corresponds to a circular velocity of $\vmax=4\,\kms$. Prior
to reionization, at a redshift of 13.6, there are 61 halos with
$\vmax(z=13.6)>4\,\kms$ in the Via Lactea simulation that will have a
surviving bound remnant today within $\rtwo$. At this early epoch
their masses range between $1.5\times10^6\,\msun$ and
$1.6\times10^8\,\msun$, and only one object in the sample is above the
critical mass threshold for atomic cooling. Let us assume that the
observed dSphs are fossils of this first generation of small-mass
galaxies, a hypothesis consistent with the fact that old populations
with ages $>10$ Gyr are the dominant stellar component of Milky Way
dSphs (see e.g. \citealt{Grebel2000,Dolphin2005}). A cosmic UV
background at $z>10$ has been shown to have little impacts on dwarf
galaxy-sized objects with $\vmax\gta 10\,\kms$ that have already grown
to a susbstantial overdensity when the ionizing radiation turns on and
can self-shield against it \citep{Dijkstra2004}. The most massive of
these early-forming objects (EF sample) may then retain a fraction of
their gas even after reionization, and may undergo a later episode of
star formation triggered e.g. by tidal shocking during their accretion
into the parent halo. By contrast, after reionization breakthrough,
photoionization heating leads to a dramatic increase in the
temperature of the intergalactic medium, and gas cannot accrete and
cool to fuel star formation in newly forming dark matter halos that
are not sufficiently massive. The exact value of this threshold has
been the subject of many studies, and depends on the redshift of
collapse, the amplitude and spectrum of the ionizing background, and
the ability of the system to self-shield against UV radiation (e.g.
\citealt{Efstathiou1992,Thoul1996,Navarro1997,Dijkstra2004}). Here we
assume that after reionization gas infall and cooling occurs only onto
dwarf galaxy-sized objects that reach the threshold
$\peakvmax>38\,\kms$. This brings our EF sample to a total of 65
surviving subhalos (as in the LBA sample). Their present-day circular
velocity function is shown in Figure \ref{velf_sat}, and it
appears to reproduce reasonably well the velocity distribution of the
known Milky Way satellites.

\begin{figure*}
\epsscale{0.85}
\plotone{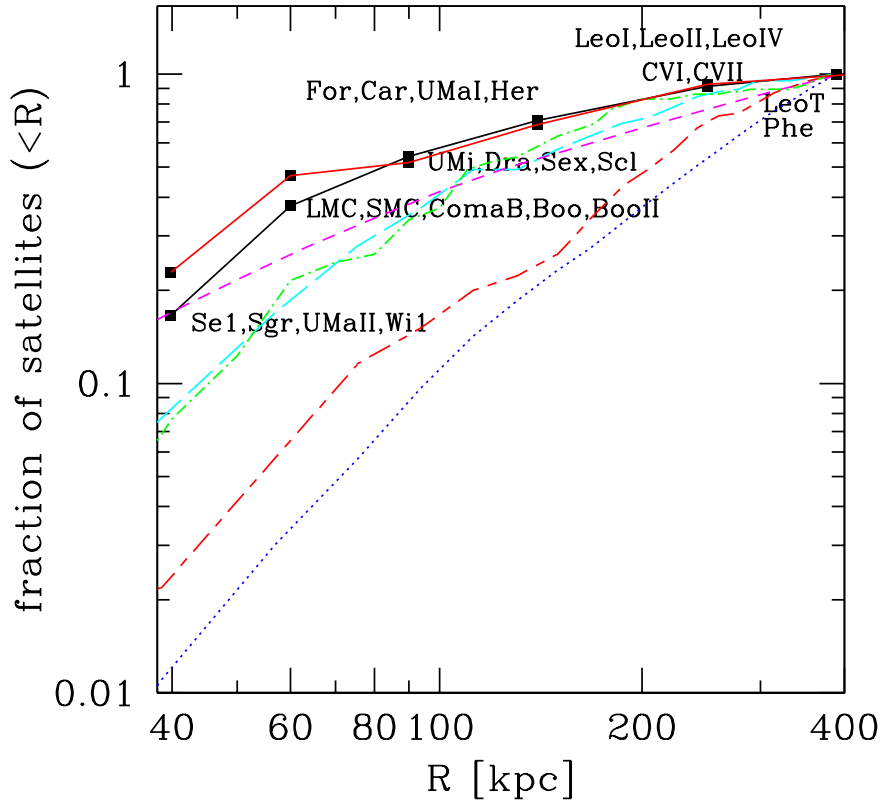}
\caption{The normalized cumulative radial distribution of Milky Way satellites ({\it lower 
solid curve}). The upper solid curve connecting the squares shows the expected 
distribution of luminous satellites after correcting for the sky coverage of the SDSS. 
For plotting and comparison purposes, we have placed Leo T and Phoenix within $\rtwo=389\,$kpc.     
{\it Long-dashed curve:} distribution for the 65 largest $\peakvmax$ subhalos before accretion 
(Via Lactea LBA sample). {\it Dot-short dashed curve:} ``fossil of reionization'' EF sample. 
{\it Short-dashed-long dashed curve:} 65 largest $\vmax$ Via Lactea subhalos today.
{\it Dotted curve:} all 4021 Via Lactea subhalos with $\msub/\mhost>10^{-6}$. 
{\it Short-dashed curve:} dark matter distribution.
}
\label{radial2}
\end{figure*}

\section{Subhalo radial distribution and central densities}

The spatial distribution of luminous satellites provides another
challenge to $\Lambda$CDM substructure models. Figure \ref{radial2}
compares the normalized cumulative radial distributions of the known
Milky Way dwarfs to that of the oldest (EF) and highest $\peakvmax$
(LBA) dark matter subhalos of Via Lactea. For plotting and comparison
purposes, we have placed Leo T and Phoenix within $\rtwo$. The
observed satellites are clearly more biased to lie at small radii than
both the LBA and the EF samples.  The median distance of the observed
satellites is 90 kpc, and the satellite distribution becomes even more
compact after correcting for the SDSS sky coverage (see Fig.
\ref{radial2}).  This must be compared to a median distance of 110 kpc
for the LBA and EF samples, and to 235 kpc for all Via Lactea
subhalos.  About 40\% (50\% after SDSS correction) of all Milky Way
dwarfs are found within 60 kpc from the Galactic center, compared to
only 20\% for the LBA and EF samples, and to 3\% for the subhalo
population as a whole. The largest subhalos today are found at
significantly larger radii than both the LBA and the EF subpopulation.
A similar discrepancy has been previously pointed out by
\citet{Kravtsov2004} and \citet{Willman2004}, and may only partially
be resolved by incompleteness in the census of distant Milky Way
dwarfs or by numerical overmerging of subhalos in the inner 50 kpc of Via
Lactea. Note that the luminous satellites appear to be biased towards
the center even compared to the overall Via Lactea's dark matter distribution.

\begin{figure*}
\plottwo{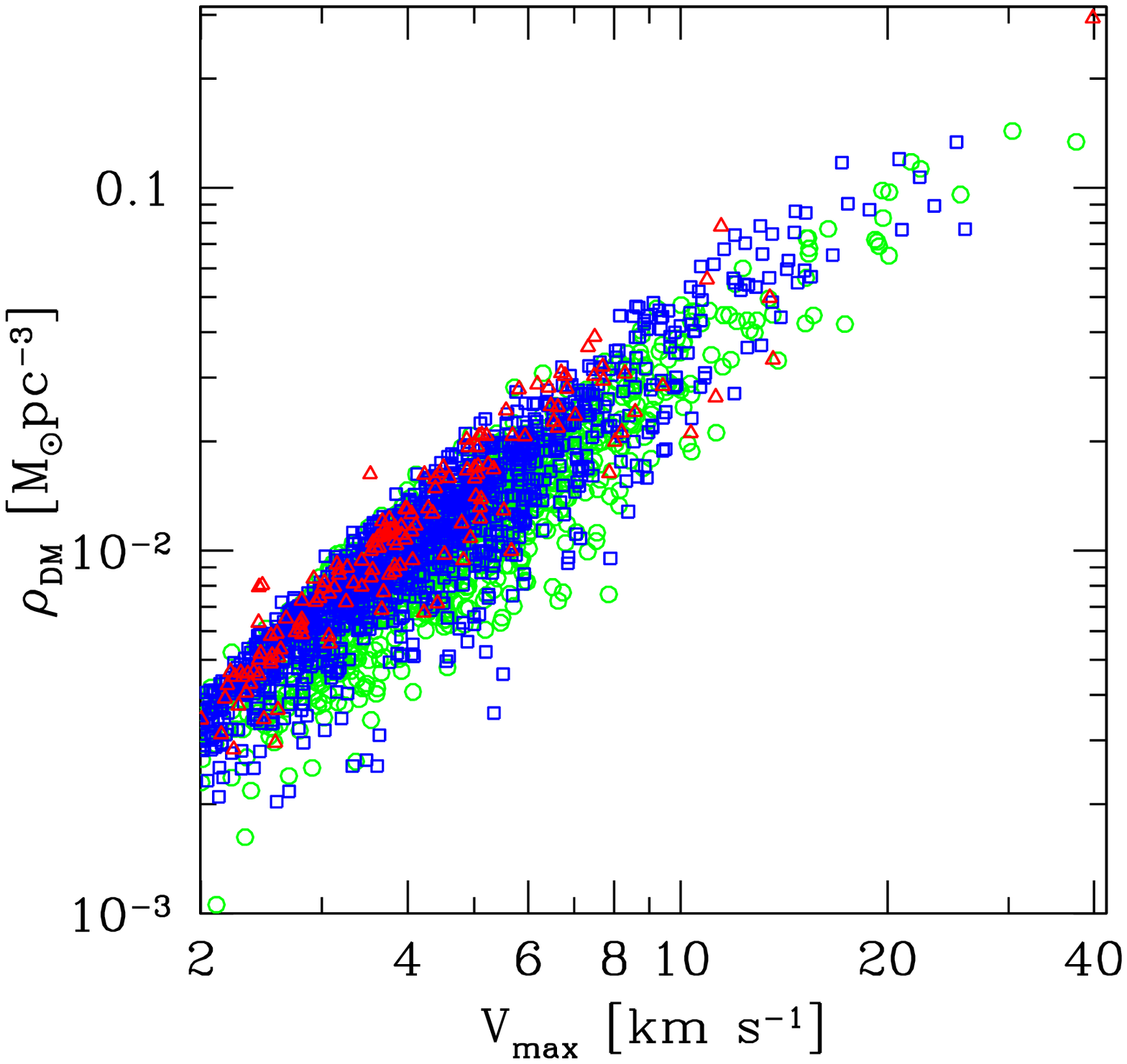}{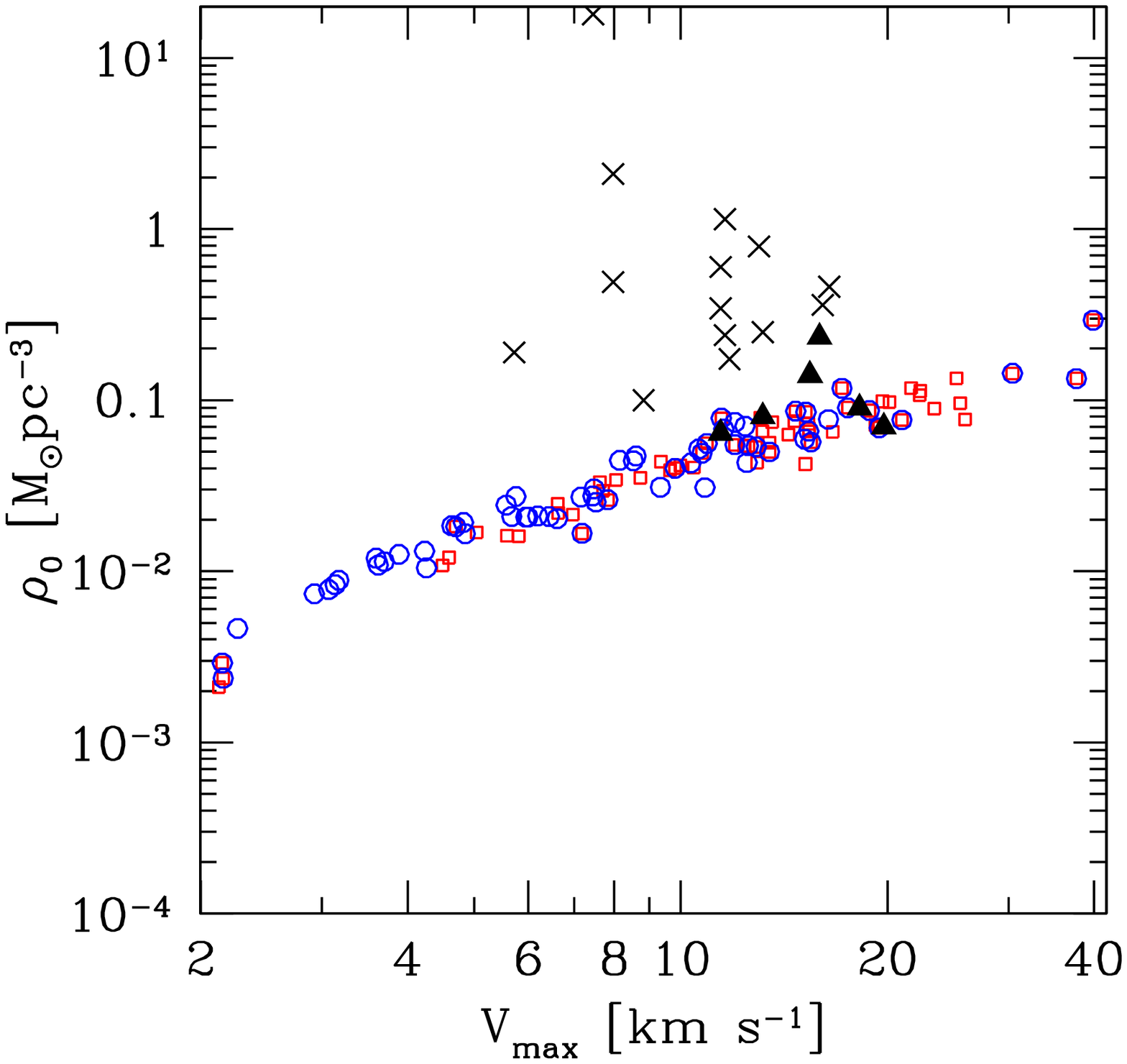}
\vspace{0.5cm}
\caption{{\it Left panel:} Central dark matter densities (measured within the inner 
300 pc) of Via Lactea subhalos as a function of their peak circular velocity today.
{\it Triangles:} all subhalos within 50 kpc from the center. 
{\it Squares:} all subhalos between 50 and 200 kpc. 
{\it Circles:} all subhalos between 200 kpc and $\rtwo$. 
{\it Right panel:} Derived central densities of Milky Way dSphs, $\rho_0=166 \sigma^2/r_c^2$, 
where $\sigma$ is the stellar velocity dispersion in $\kms$ and $r_c$ is the King
profile core radius in pc (see text for details). {\it Diagonal crosses:} dSphs with 
$r_c<250$ pc.   
{\it Filled traingles:} dSphs with $r_c>250$ pc.    
{\it Empty circles:} EF subhalo sample. 
{\it Empty squares:} LBA subhalo sample. 
}
\label{rho}
\end{figure*}

The study of the central densities of dark-matter dominated dSphs
offers another test of $\Lambda$CDM galaxy formation models and of the
systematic small-scale properties -- only minimally modified by
baryonic effects -- of their halos. To estimate the central mass
density, $\rho_0$, of pressure-supported low luminosity dwarfs, a
``core fitting'' or King's analysis is usually performed
\citep{King1966,Richstone1986}. Under the assumption of spherical
symmetry, close to dynamical equilibrium, mass following light, and
isotropic stellar velocity dispersion, one has \be
\rho_0={166\eta\sigma^2\over r_c^2}\,\sden,
\label{density}
\ee where $\sigma$ is the observed central velocity dispersion in
$\kms$, $r_c$ the King profile core radius in pc, and $\eta$ is a
numerical parameter very close to unity. In two-component isotropic
models in which luminous matter is negligible compared to dark matter
and in which the core radius of the dark matter is larger than that of
the visible material, equation (\ref{density}) can overestimate the
true central density by up to a factor of 2.2 \citep{Pryor1990}. We
have evaluated the central densities of Milky Way's spheroidals from
equation (\ref{density}) (setting $\eta=1$) using data in the
literature. For the ultra-faint new dwarfs, King core radii have been
estimated from published Plummer (half-light) radii $r_P$ using
$r_c=0.64\,r_P$ \citep{Simon2007}. The radii and velocity dispersions
assumed for this calculation, as well as our derived central
densities, are given in Table 1. Central densities range from $\lta
0.1\,\sden$ (e.g. Hercules, Sextans, Canes Venatici I) to $\sim
1\,\sden$ (Ursa Major I, Coma Berenices, Leo T) to $\sim 20\,\sden$
(Willman 1). For comparison, we have computed the mean dark matter
densities within 300 pc (about 3 times our force resolution) from
their center, $\rho_{\rm DM}$, for all Via Lactea's subhalos.  These
are plotted in the left panel of Figure \ref{rho} as a function of
peak circular velocity for subhalos at distances $R<50$ kpc,
$50<R<200$ kpc, and $200<R<\rtwo$.  These central densities increase
steeply with circular velocities. In the smaller systems a distance of
300 pc is comparable to the radius where the peak circular velocity is reached,
i.e. we probe scales where halo
profiles are nearly isothermal and indeed the small systems follow the
$\rho_{\rm DM} (<300 {\rm pc}) \propto \vmax^2$ scaling expected for
isothermal halos quite well.  The relation flattens for larger
subhalos since their scale radii are larger than 300 pc and our
measuments now probe the inner regions, which are shallower than
isothermal.

The right panel of Figure \ref{rho} shows a comparison between the
dark matter densities of subhalos in the LBA and EF sample and the
central densities of dSphs.  Via Lactea is clearly starting to resolve
the inner dense regions of Milky Way-like satellites. In particular,
the central densities of dSphs with measured core radii in excess of
250 pc (i.e. comparable to the radii within which $\rho_{\rm DM}$ is
measured) appear to be in good agreement with the values found in
subhalos of comparable circular velocities. A few satellites with
core radii smaller than 100 pc, like Ursa Major II, Coma Berenices, and
Willman 1, have central densities in excess of $1\,\sden$. The predicted 
dark matter densities we obtain by extrapolating inward the steep inner 
profiles of subhalos are perfectly consistent with the observed values. 
The left panel of the same
figure also demonstrates that the typical central densities at a given
$\vmax$ are systematically larger for inner subhalos. This is a
consequence of the larger average tidal mass loss suffered by this
subpopulation: tidal stripping can significantly reduce $\vmax$
without changing the central density, and inner subhalos end up having
much larger typical concentrations \citep{Diemand2007b}.  This
concentration-radius relation should be taken into account when the
central densities of luminous dwarfs (constrained by stellar
kinematics) are used to infer the peak circular velocities of the host
subhalos.

A simple way for estimating the range of host subhalo sizes
corresponding to a given Milky Way satellite is to use the integrated
dSph ``central mass'' within a physical radius comparable to the
extent of the luminous region, and compare it to those of Via Lactea's
subhalos at similar galactocentric distances. The central masses
$M_{0.6}$ (mass within 0.6 kpc) and $M_{0.1}$ (mass within 0.1 kpc)
are robustly constrained by kinematic data. Given Via Lactea's force
resolution of 90 pc, $M_{0.6}$ is also robustly determined in
simulated subhalos. $M_{0.1}$, however, is likely affected by
gravitational softening and may be systematically underestimated. Here
we use the most likely values and 90\% confidence regions of $M_{0.1}$
from \citet{Strigari2007b} and of $M_{0.6}$ from
\citet{Strigari2007a}, as well as the best-fit $M_{0.6}$ values from
\citet{Walker2007}. We then find all Via Lactea subhalos within 40\%
in distance and within 25\% of the $M_{0.1}$ or $M_{0.6}$ values of
each satellite, and then average over their $\vmax$ and $\peakvmax$ to
find the most likely host size.\footnote{For the dSphs marked with an
  asterisk in Table \ref{sizetable} no subhalos were found in the
  quoted range, and an estimate of the most likely host size was made
  using the subhalo with the smallest quadratic sum of the relative
  difference in distance and central mass.}\ The range of plausible
peak circular velocities was determined using all subhalos within the
90\% confidence region in $M_{0.1}$ (or $M_{0.6}$).  The results of
this exercise are given in Table \ref{sizetable}. Consistent with the
expectations from the concentration-radius relation, for a given dSph
central mass the most likely $\vmax$ values of its host subhalos are
larger at large galactocentric distances. If, e.g. Ursa Minor with
$M_{0.6}=5.3\,\msun$ were to lie at a distance of 300 kpc, its most
likely host would have $\peakvmax=31\,\kms$ and $\vmax=26\,\kms$.
Coma Berenices with $M_{0.1}=1.9\,\msun$ at 300 kpc would typically be
hosted by a $\peakvmax=30\,\kms$ and $\vmax=21\,\kms$ subhalo.

\begin{deluxetable*}{l c c c c c l}
\tablewidth{0pt}
\tablecolumns{6}
\tablecaption{Likely host subhalos of Milky Way satellites}
\tablehead{
\colhead{Name} & \colhead{Distance} & \colhead{$\sigma$} 
& \colhead{Central mass$^a$} 
& \colhead{$\peakvmax$} & \colhead{$\vmax$} & \colhead{References}\\
\colhead{} & \colhead{kpc} & \colhead{$\kms$} & \colhead{$10^7 \msun$} &
\colhead{$\kms$} & \colhead{$\kms$} & \colhead{}
}

\startdata

Ursa Major II$^*$ & 38 & 6.7 & $0.31^{+0.56}_{-0.1}$ & 50[15-67]&11[11-21] & (3)\\
Willman 1       & 43 & 4.3 & $0.13^{+0.15}_{-0.08}$ & 37 [ 9-50]&14 [ 5-17] & (3)\\
Coma B          & 45 & 4.6 & $0.19^{+0.21}_{-0.1}$ & 41 [14-63]&16 [ 8-18] & (3)\\
Ursa Minor      & 66 & 9.3 & $0.23^{+0.19}_{-0.12}$ & 28 [14-63]&17 [ 8-21] & (3)\\
\hline
Sagittarius$^*$ & 16 & 11.4& $27^{+20}_{-27}$ & 94 & 40 & (1) \\
Ursa Minor      & 66 & 9.3 & $5.3^{+1.3}_{-1.3}$ & 49 [35-63] &20 [17-22] & (1)\\
Draco           & 79 & 9.5 & $4.9^{+1.4}_{-1.3}$ & 51 [30-67] & 21 [21-23] & (1)\\
Draco$^*$       & 79 & 9.5 & 6.9 & 35 & 22 & (2)\\
Sextans         & 82 & 6.6 & $0.9^{+0.4}_{-0.3}$ & 14 [ 8-23] & 7 [ 6-9] & (1)\\
Sextans         & 82 & 6.6 & 2.5 & 25 & 14 & (2) \\
Sculptor	& 86 & 6.6 & $2.7^{+0.4}_{-0.4}$ & 26 [21-29] & 15 [12-17] & (1) \\
Sculptor	& 86 & 6.6 & 4.3 & 32 & 23 & (2)\\
Carina          &101 & 6.8 & $3.4^{+0.7}_{-1.0}$ & 29 [28-30]& 18 [14-23] & (1) \\
Carina          &101 & 6.8 & 2.0 & 25 & 12 & (2) \\
Fornax          &138 & 10.5& $4.3^{+2.7}_{-1.1}$ & 42 [21-63]& 23 [15-26] & (1) \\
Leo II          &205 & 6.6 & $2.1^{+1.6}_{-1.1}$ & 21 [9-33]& 13 [8-18] & (1) \\
Leo I           &250 & 8.8 & $4.3^{+1.6}_{-1.6}$ & 29 [21-35] & 23 [15-26] & (1)
\enddata

\tablenotetext{a}{Equal to $M_{0.1}$ in the top 4 systems and to $M_{0.6}$ in all 
others. The quoted ranges are 90\% confidence level regions.} 
\tablenotetext{*}{For these dSphs no subhalos were found in the quoted range, and an 
estimate of the most likely host size was made us described in the text.}
\tablerefs{(1) \citet{Strigari2007a}; (2) \citet{Walker2007}; (3) 
\citet{Strigari2007b}.
}
\label{sizetable}
\end{deluxetable*}

\section{Summary and conclusions}

In this paper we have reported results from the two highest resolution
simulations of Galactic CDM substructure to date, Via Lactea and
1e8Ell. The two runs follow the hierarchical assembly of a Milky
Way-sized halo down to $z=0$ and of an elliptical-sized halo down to
$z=0.5$, respectively, using different cosmological parameters. Our
main findings can be summarized as follows:

\begin{enumerate} 

\item Via Lactea and 1e8Ell are characterized by similarly steeply
  rising subhalo mass functions.  In the range
  $200m_p<\msub<0.01\mhost$, the best-fit slope of the differential
  distribution, $dN/d\msub\propto \msub^{\alpha}$, is $\alpha=-1.86\pm
  0.02$ for 1e8Ell and $\alpha=-1.90\pm 0.02$ for Via Lactea. In the
  same mass range the cumulative mass function has slope $-0.92$ for
  1e8Ell and $-0.97$ for Via Lactea.  Compared to Via Lactea, 1e8Ell
  produces nearly a factor of two more subhalos with large circular
  velocities. When normalized to the maximum circular velocity of the
  host, the substructure circular velocity distribution yields 170
  subhalos above $\vmax=22.5\, \kms=0.056\,\vhost$ in 1e8Ell, compared
  to 110 above $\vmax=10\,\kms=0.056\,\vhost$ in Via Lactea;

\item the fraction of the Via Lactea's host halo mass brought in by
  subhalos that have a surviving bound remnant today within $\rtwo$
  and with peak circular velocity $\vmax>2\,\kms$ is 45\%. 
  The total mass accreted in identifiable subunits exceeds 97\%, 
  as many major mergers at early times left no surviving $z=0$ subhalos. 
  Hence at our high resolution most of the Via Lactea mass is acquired in 
  discrete clumps, with no significant contribution from smooth 
  accretion.  
  There are $(14,109,873)$ subhalos today
  with $\vmax>(20,10,5)\,\kms$, which contributed $(19,30,39)$\% of
  the mass of Via Lactea. Only 9-10\% of the total material they brought in
  remains in self-bound surviviving lumps at the present-epoch, the
  rest having been removed by tidal effects to become part of the
  ``smooth'' halo component. Mass losses are more severe for larger
  subunits. 

\item because of tidal mass loss, the number of Via Lactea subhalos
  that have a surviving bound remnant today and reached a peak
  circular velocity of $>10\,\kms$ throughout their lifetime exceeds
  half a thousand, five times larger than their present-day abundance
  and more than twenty times larger than the current (incomplete)
  tally of observed dwarf satellites of the Milky. The 20 largest
  $\vmax$ subhalos today have lost between 30\% and 99.5\% of the
  maximum mass that had before accretion. If substructure mass regulates star
  formation, then for a given mass threshold many more subhalos should
  have been able to build a sizeable stellar mass at some point in the
  past than indicated by their present-day abundance.

\item unless the circular velocity profiles of Galactic satellites
  peak at values significantly higher that expected from the stellar
  line-of-sight velocity dispersion, only one out of five subhalos
  with $\vmax>20\,\kms$ today must be housing a luminous Milky Way
  dwarf. Any mechanism that suppresses star formation in small halos
  must start acting early, at redshift $z>3$.  Nearly 600 halos with
  masses greater than $10^7\,\msun$ are found today in the ``field''
  between $\rtwo$ and $1.5\rtwo$, i.e. small dark matter clumps appear
  to be relatively inefficient at forming stars even well beyond the
  virial radius. Even if most subhalos appear to have no optically
  luminous counterparts, such a large substructure population may be
  detectable via flux ratio anomalies in strong gravitational lenses
  (e.g. \citealt{Metcalf2001}), through its effects on stellar streams
  (e.g. \citealt{Ibata2002,Mayer2002}), or possibly via $\gamma$-rays
  from dark matter annihilation in their cores
  \citep{Bergstrom1999,Stoehr2003};

\item the observed Milky Way satellites approximately follow the
  overall dark matter distribution of Via Lactea. They appear to be
  more biased to lie at small radii than both the LBA and the EF
  samples.  The median distance of the observed satellites is 90 kpc,
  and the satellite distribution becomes even more compact after
  correcting for the SDSS sky coverage. This must be compared to a
  median distance of 110 kpc for the LBA and EF samples, and to 235
  kpc for all Via Lactea subhalos.  About 40\% (50\% after SDSS
  correction) of all Milky Way dwarfs are found within 60 kpc from the
  Galactic center, compared to only 20\% for the LBA and EF samples,
  and to 3\% for the subhalo population as a whole. Such discrepancies
  may only be partially resolved by incompleteness in the census of
  distant Milky Way dwarfs or by numerical overmerging in the inner 
  50 kpc of Via Lactea.

\item the study of the central densities of dark-matter dominated
  dSphs offers another test of $\Lambda$CDM galaxy formation models
  and of the systematic small-scale properties -- only minimally
  modified by baryonic effects -- of their halos. Via Lactea subhalos
  have central matter densities that increase with $\vmax$ and reach
  $\rho_{\rm DM}=0.1-0.3\,\sden$, comparable to the central densities
  inferred in Milky Way dwarfs with core radii $>250$ pc.

\end{enumerate}

\acknowledgments P.M. acknowledges support from NASA grant NNG04GK85G.
J. D. acknowledges financial support from NASA through Hubble
Fellowship grant HST-HF-01194.01 awarded by the Space Telescope
Science Institute, which is operated by the Association of
Universities for Research in Astronomy, Inc., for NASA, under contract
NAS 5-26555. M.K. gratefully acknowledges support from the Institute
for Advanced Study. All computations were performed on NASA's Project
Columbia supercomputer system.

{}

\end{document}